\documentclass[aps,groupedaddress]{revtex4}
\usepackage{epsfig}
\usepackage{graphicx}
\usepackage[T1]{fontenc}
\usepackage{ae}
\usepackage{ulem}
\usepackage{color}
\usepackage[latin1]{inputenc}
\usepackage{amssymb,amsbsy,amsmath}
\usepackage{bbm}

\newtheorem{lemma}{Lemma}
\newtheorem{defi}{Definition}
\newtheorem{prop}{Proposition}

\newtheorem{theorem}{Theorem}

\input ulem.sty

\begin{document}


\title[GCI]
      {A generalization of the Choi isomorphism with application to open quantum systems}
\author{Heinz-J\"urgen Schmidt$^1$}
\address{$^1$  Universit\"at Osnabr\"uck,
Fachbereich Mathematik, Informatik und Physik,
 D - 49069 Osnabr\"uck, Germany}


\begin{abstract}
Completely positive transformations play an important role in the description of state changes
in quantum mechanics, including the time evolution of open quantum systems.
One useful tool to describe them is the so-called Choi isomorphism,
which maps completely positive transformations to positive semi-definite matrices.
Accordingly, there are numerous proposals to generalize the Choi isomorphism.
In the present paper, we show that the 1976 paper of Gorini, Kossakowski and Sudarshan (GKS)
already holds the key for a further generalization and study the resulting GKS isomorphism.
As an application, we compute the GKS matrix of the time evolution of a
general open quantum system up to second order in time.
\end{abstract}

\maketitle
\section{Introduction}\label{sec:Intro}

It has proved useful to employ a general concept of ``state changes" in quantum mechanics, which
includes time evolutions of closed and open systems, as well as the changes due to measurements which
generalize von Neumann's projection postulate. Since (pure or mixed) quantum states
are represented by statistical operators (density matrices),
it is natural to describe such state changes by linear mappings that preserve the trace
and map positive operators to positive ones.

Less obvious is the additional postulate that state changes should be ``completely positive".
This property can be roughly described by saying that it should always be possible to extend
a state change in a system $S$ to a larger system $S+E$ in a way
that it acts on $E$ like the identity. It is then plausible that, e.~g.,
the restriction of a unitary time evolution in $S+E$ to the system $S$
by taking the partial trace leads to a completely positive mapping.
More generally, a state change of $S+E$ caused by a von Neumann measurement can be reduced to
to a completely positive state change of $S$.

From the mathematical point of view, ``complete positivity" was first defined in the paper of
Stinespring \cite{S55}, who considers maps between $C^\ast$-algebras, whose duals are the above mentioned
state transformations. This concept of a completely positive map
has been taken up in the mathematical literature and further developed in the work of
St{\o}rmer \cite{S63} and Choi \cite{C75}, for example. The latter has given
a simplified criterion for complete positivity, which will be generalized in the present work.
It is also worthwhile to mention the paper of Jamio{\l}kowsi \cite{J72} which completes a
previous paper of de Pillis \cite{dP67} who defines what is known as the ``Jamio{\l}kowsi isomorphism".
It has some similarities with Choi isomorphism, but, as recently pointed out by
Frembs and Cavalcanti \cite{FC24}, it is different and does not provide a criterion for complete positivity,
see also subsection \ref{sec:COG1} of the present paper.

In physics, the realization that the general changes of state in quantum mechanics
should be described by completely positive mappings only became established over a longer process of about a decade.
In retrospect, the work of Sudarshan et al \cite{SMR61} can be seen as a forerunner
in which this insight begins to emerge. Let us therefore take a closer look at the details of the argumentation.
In the context of open quantum systems the authors of \cite{SMR61} consider finite-dimensional
density matrices $\left(\rho_{rs}\right)$ and linear maps of these given by matrices $\left(A_{rs,r's'}\right)$.
From the condition (12) equivalent to $\rho\ge 0$ the authors draw the ``consequence" (12') which is a condition
on the matrix $A$ equivalent to complete positivity. Obviously, the authors of \cite{SMR61} did not
know the work of Stinespring \cite{S55} and hence did not use the term ``completely positive".
Also, from today's point of view, one would  say that (12') is not a consequence,
but rather a sufficient condition for $A$ mapping positive matrices to positive matrices.
Interestingly, the authors of \cite{SMR61} then define in (14) a matrix $B$ by $B_{r r', s s'}=A_{rs, r's'}$
and formulate the (easy to prove) statement that (12') is equivalent to $B\ge 0$.
This is already in essence the Choi isomorphism and the associated criterion for complete positivity.

In the years that followed, the term ``operation" was introduced for certain state changes
in the work of Haag and Kastler \cite{HK64} and further elaborated by
Davies and Lewis \cite{DL70}. The first steps towards a restriction of the general concept of an
``operation" were taken by Hellwig and Kraus in \cite{HK69} and \cite{HK70},
who developed their approach under the influence of Ludwig's axiomatic foundation of quantum mechanics \cite{L70}.
They derived what is known today as the ``operator-sum representation" or ``Kraus operator representation"  of an operation,
see, e.g., \cite{NC00},
and which has been recognized as being equivalent to complete positivity in \cite{K71}.
These results are usually quoted from the ``Lecture notes in physics" volume \cite{K83}.

After this breakthrough, it is not surprising that, as already alluded to in \cite{SMR61},
completely positive mappings are also used in the time evolution of open quantum systems.
In 1976 two papers appeared simultaneously investigating semigroups of completely positive time evolutions,
\cite{GKS76} and \cite{L76}, which led to the so-called Lindblad equation.
This term has become commonplace, though perhaps not entirely fair.
It is interesting for the history of the Lindblad equation that,
acording to Vittorio Gorini \cite{VG24},
the first author of \cite{GKS76} became acquainted with the concept of
completely positive mappings in discussions with K.~Kraus
at a conference in Marburg in 1973 \cite{HN74}.
He told this to A.~Kossakowski and informed E.~Sudarshan,
who invited to Austin both Kossakowski and Gorini for a few months in 1975-76,
where the paper \cite{GKS76} was written.
This part of the history of the Gorini-Kossakowski-Sudarshan-Lindblad equation is also touched in 
\cite{CP17} and in \cite{W25}.

For our purposes, eq.~(2.4) in \cite{GKS76} is particularly interesting,
since it establishes a 1:1 relationship between linear maps ${\mathcal E}$ of state spaces and Hermitean matrices,
called GKS matrices $g$ in this paper, independent of the semigroup property.
Consequently, the approach of \cite{GKS76} was later used to derive a
time-dependent Lindblad equation which holds for general open quantum systems,
see \cite{B12} or subsection \ref{sec:TDLE} of this paper.
We will prove that ${\mathcal E}$ is completely positive iff $g$ is positively semi-definite,
thus generalizing Choi's criterion, since the Choi matrix can be seen as a special case of the GKS matrix.
This result is already partially included in papers on ```quantum process tomography"; 
see eq.~(6) in \cite{CN97}, eq.~(8.152) in \cite{NC00} and \cite{Al03}, where the GKS matrix appears under the name ``chi matrix."

In quantum information theory, ``quantum channels" are communication channels
that can carry quantum information as well as classical information.
Mathematically, they are defined as completely positive trace-preserving maps between
spaces of operators, and thus as ``operations" in the sense mentioned above.
This definition can be traced back to \cite{H72} and has been therefore developed at the same time
as the other physical applications mentioned above.
The Choi isomorphism is also known as ``channel-state duality" in quantum information theory.
It can be used, for example, to  parameterize completely positive trace preserving  maps for optimizing quantum processes,
see \cite{HOK24}.

The paper is organized as follows. After establishing some basic definitions and general results in
Section \ref{sec:Def}, we recapitulate the Choi isomorphism and its properties in
Section \ref{sec:CI}, closely following \cite{FC24}.
The GKS isomorphism is described in section \ref{sec:GKSI}. We use the notation of ${\mathcal A}$
to denote the Hilbert-Schmidt class of linear operators on some $N$-dimensional Hilbert space.
The GKS-matrix of a general linear superoperator ${\mathcal E}: {\mathcal A} \rightarrow {\mathcal A}$
will be defined w.~r.~t.~a general orthonormal basis in ${\mathcal A}$. For the special choice
of the orthonormal basis with elements $|i\rangle \langle j|$ the GKS-matrix reduces to the
Choi matrix. This will be shown in subsection \ref{sec:GKSCI}.
As a result, the GKS matrix inherits some favorable properties from the Choi matrix,
which allows us to formulate the final definition in the subsection \ref{sec:DGKSI}
and its properties in \ref{sec:PGKSI},
the most important being the criterion for complete positivity of ${\mathcal E}$.
In Section \ref{sec:COI} we compare the GKS isomorphism with other generalizations
of the Choi isomorphism that have been proposed recently and show that it differs from them.

As already mentioned, the GKS matrix first appeared in the context of open quantum systems.
Therefore, it will be appropriate to take a closer look at its connection with
the time-dependent Lindblad equation and the role of the criterion of
complete positivity for the time evolution of general open quantum systems.
This will be done in section \ref{sec:A}. First we consider the general form
of the differential equation for a time-dependent superoperator ${\mathcal E}(t)$,
called GKS equation, and transform it into a differential equation for the GKS matrix $g(t)$,
see subsection \ref{sec:TDGKS}. Then, in subsection \ref{sec:TDLE}, we recapitulate the derivation
of the time-dependent Lindblad equation starting from the GKS equation and, additionally, consider
the reverse derivation.

These calculations do not yet assume that the GKS equation follows from the reduced interaction
of the system $S$ with an environment $E$. In the subsection \ref{sec:EXP}, however, we make this assumption
and try to calculate the GKS matrix as a function of the Hamiltonian for the total system $S+E$.
Due to the inherent difficulty of these calculations, we restrict ourselves to a series expansion
up to the $2$nd order w.~r.~t.~time $t$. Note that in the dynamic semigroup approximation,
the reduced time derivative of the statistical operator $\rho(t)$ would be constant
and thus its $2$nd derivative would vanish.
Our calculations therefore lead to results that go beyond the semigroup approximation.
We will focus on the question of whether the GKS matrix is positively semi-definite
in $2$nd order in $t$. On the one hand,
this follows in general since the time evolution of $\rho(t)$ is completely positive.
On the other hand, it is instructive to show this directly in order to check the
consistency of the present approach.  We conclude with a summary in Section \ref{sec:SUM}.

\section{Definitions and general results}\label{sec:Def}

We put together some well-known definitions to fix our notation.
Let ${\mathcal H}$ be an $N$-dimensional complex Hilbert space and ${\mathcal A}:=L({\mathcal H})$ denote the complex
$N^2$-dimensional Hilbert space of linear operators $A:{\mathcal H} \rightarrow {\mathcal H}$
(``Hilbert-Schmidt class")
endowed with the scalar product
\begin{equation}\label{defscalarproductA}
 \langle X\left|\right.Y\rangle:= \mbox{Tr }\left(X^\ast \, Y\right),\;
X,Y\in {\mathcal A}
 \;,
\end{equation}
where $\mbox{Tr}$ denotes the trace and $X^\ast$ the adjoint of the operator $X$.
Further let ${\sf B}({\mathcal H})$ denote the real subspace of $L({\mathcal H})$ consisting of Hermitean operators,
${\sf B}_0({\mathcal H})\subset {\sf B}({\mathcal H})$ the subset of operators with vanishing trace,
${\sf B}^+({\mathcal H})\subset {\sf B}({\mathcal H})$ the cone consisting of positively semi-definite operators,
${\sf B}^+_1({\mathcal H})\subset {\sf B}^+({\mathcal H})$ the convex set of statistical operators.
and ${\sf U}({\mathcal H})$ the group of unitary operators acting on ${\mathcal H}$.
There exists a canonical isomorphism $\imath: {\mathcal A} \otimes {\mathcal A}  \rightarrow L({\mathcal H}\otimes {\mathcal H})$
given by
\begin{equation}\label{identAA}
 \imath (A\otimes B)(\varphi\otimes \psi):=A(\varphi)\otimes B(\psi),\quad A,B\in {\mathcal A},\; \varphi,\psi\in{\mathcal H}
 \;,
\end{equation}
and linear extension w.~r.~t.~operators and vectors.
We will henceforward identify ${\mathcal A} \otimes {\mathcal A} $ and $L({\mathcal H}\otimes {\mathcal H})$.

Let ${\mathcal H}_i,\,i=1,2$ be two finite-dimensional Hilbert spaces.
Then the  partial trace $\mbox{Tr}_2: L({\mathcal H}_1\otimes {\mathcal H}_2) \rightarrow L({\mathcal H}_{1})$
is defined by
\begin{equation}\label{defpartialtrace}
\mbox{Tr}_2(A\otimes B)= A\,  \mbox{Tr }(B),\quad A\in L({\mathcal H}_1),\;B\in L({\mathcal H}_2)
 \;,
\end{equation}
and linear extension. Analogously for  $\mbox{Tr}_1: L({\mathcal H}_1\otimes {\mathcal H}_2) \rightarrow L({\mathcal H}_{2})$.

After choosing an orthonormal basis $(| i\rangle)_{0\le i <N}$ in ${\mathcal H}$ we may identify ${\mathcal H}$ with
${\mathbb C}^N$. To simplify the notation, $L({\mathbb C}^N)$ will be identified with the space of complex $N\times N$-matrices
and analogously for ${\sf B}({\mathbb C}^N)$, ${\sf B}^+({\mathbb C}^N)$, and ${\sf U}({\mathbb C}^N)$.

Since ${\mathcal A}$ is an $N^2$-dimensional Hilbert space the above definitions can be iterated and yield
$L({\mathcal A})$ as a complex $N^2\times N^2$-dimensional Hilbert space of ``superoperators" w.~r.~t.~the scalar product
\begin{equation}\label{defscalarproductLA}
 \langle {\mathcal X}\left|\right.{\mathcal Y}\rangle := \mbox{Tr} \left( {\mathcal X}^\ast \, {\mathcal Y}\right),\;
 {\mathcal X},{\mathcal Y}\in L({\mathcal A})
 \;.
\end{equation}
After introducing an orthonormal basis in ${\mathcal A}$, the space $L({\mathcal A})$ can be identified with
the space $L({\mathbb C}^{N\times N})$ of complex $N^2\times N^2$-matrices.

We will formulate the definitions concerning ``complete positivity" only for the special case
of a linear map $\phi: {\mathcal A}\rightarrow{\mathcal A}$ considered in this article, following \cite{NC00}
and \cite{C06}. Recall that $L\left({\mathbb C}^M\right)$ denotes the algebra of complex $M\times M$-matrices.
$\phi$ can be extended to $\phi\otimes {\mathbbm 1}_N:{\mathcal A}\otimes L\left({\mathbb C}^M\right) \rightarrow {\mathcal A}\otimes L\left({\mathbb C}^M\right)$
by $(\phi\otimes {\mathbbm 1}) (A\otimes M):=\phi(A)\otimes M$ and linear extension.
Then $\phi$ is called ``completely positive" iff $X\ge 0$ implies $(\phi\otimes {\mathbbm 1}_N)(X)\ge 0$ for all Hermitean operators
$X\in {\mathcal A}\otimes L\left({\mathbb C}^M\right)$.
This condition is stronger than the property that $\phi$ maps positively semi-definite operators onto  positively semi-definite ones.
The usual example to illustrate this fact is $\tau(A)=A^\top$ (transposition). $\tau$ leaves the eigenvalues of any positively semi-definite
operator invariant but is not completely positive:  Set $N=2$ and  ${\mathcal A}=L\left({\mathbb C}^2\right)$
and consider the matrix $X\in {\mathcal A}\otimes L\left({\mathbb C}^2\right)$ given by
\begin{equation}\label{X}
 X=\left(\begin{array}{cc|cc}
 1 & 0 & 0 & 1 \\
 0 & 0 & 0 & 0 \\
 \hline
 0 & 0 & 0 & 0 \\
 1 & 0 & 0 & 1 \\
\end{array}
\right)
\end{equation}
with eigenvalues $(2,0,0,0)$.
$\tau\otimes {\mathbbm 1}_2$ swaps the two anti-diagonal blocks of $X$
\begin{equation}\label{phiX}
 (\tau\otimes {\mathbbm 1}_2)(X)=\left(\begin{array}{cc|cc}
 1 & 0 & 0 & 0 \\
 0 & 0 & 1 & 0 \\
 \hline
 0 & 1 & 0 & 0 \\
 0 & 0 & 0 & 1 \\
\end{array}
\right)
\;,
\end{equation}
and thus yields a matrix with eigenvalues $(-1,1,1,1)$.

Completely positive maps are characterized by a finite (Kraus) operator sum representation:
\begin{prop}\label{PKraus}
  \begin{enumerate}
    \item Every map ${\mathcal E}:{\mathcal A}\rightarrow {\mathcal A}$ of the form ${\mathcal E}(X)=\sum_{i=1}^{n}A_i\,X\,A_i^\ast$
    with linear operators $A_i\in{\mathcal A},\,i=1,\ldots,n$ is linear and completely positive.
    \item  If ${\mathcal E}:{\mathcal A}\rightarrow {\mathcal A}$ is linear and completely positive, then it admits an operator sum
    representation ${\mathcal E}(X)=\sum_{i=1}^{n}A_i\,X\,A_i^\ast$ such that $n\le N^2$ and the $A_i\in{\mathcal A}$ are pairwise
    orthogonal.
  \end{enumerate}
\end{prop}
For the proof see \cite{NC00}. If  ${\mathcal E}:{\mathcal A}\rightarrow {\mathcal A}$ is completely positive with the
Kraus operator representation according to Proposition \ref{PKraus} then its adjoint
${\mathcal E}^\ast:{\mathcal A}\rightarrow {\mathcal A}$ will have the Kraus operator representation
${\mathcal E}^\ast(X)=\sum_{i=1}^{n}A_i^\ast\,X\,A_i$ and is hence also completely positive.
For the finite-dimensional case, it is therefore not necessary to distinguish between
${\mathcal E}$ and ${\mathcal E}^\ast$ w.~r.~t.~complete positivity.
Note, however, that if a completely positive map ${\mathcal E}$ is
physically understood as a transformation of states, then it is natural to postulate that ${\mathcal E}$
will be trace-preserving. This is equivalent to  ${\mathcal E}^\ast({\mathbbm 1}_N)={\mathbbm 1}_N$, where
 ${\mathcal E}^\ast$ is physically understood to operate on observables (or ``effects" in the sense of \cite{L70}).

\section{The Choi isomorphism}\label{sec:CI}

The Choi isomorphism is a linear isomorphism
$C: L({\mathcal A}) \rightarrow {\mathcal A} \otimes {\mathcal A} \cong L({\mathcal H}\otimes {\mathcal H})$
depending on an orthonormal basis $\left(|i\rangle\right)_{0\le i<N}$ in ${\mathcal H}$.
It is defined as follows. Let  $\Phi$ denote the maximally entangled state
\begin{equation}\label{defent}
 |\Phi\rangle := \sum_{i=0}^{N-1}|i\rangle\otimes |i\rangle \in {\mathcal H}\otimes {\mathcal H}
 \;,
\end{equation}
then
\begin{equation}\label{defChoi}
 C({\mathcal E}):=\left({\mathbbm 1}\otimes {\mathcal E} \right)\left( | \Phi\rangle\, \langle \Phi|  \right)
 \;,
\end{equation}
see, e.~g.~\cite{NC00}. Following \cite{FC24} we have omitted the normalization factor $1/\sqrt{N}$ in (\ref{defent}) in order to
simplify the representation. The most important properties of $C$ are summarized in the following
\begin{theorem}\label{T1}
  \begin{enumerate}
    \item $C: L({\mathcal A}) \rightarrow {\mathcal A} \otimes {\mathcal A}$ is a linear isomorphism with inverse
    \begin{equation}\label{Cinverse}
      C^{-1}:{\mathcal A} \otimes {\mathcal A} \rightarrow L({\mathcal A}),\quad
      C^{-1}({\mathcal X})(a):= \mbox{Tr}_1 \left( {\mathcal X}\left(a^\top\otimes {\mathbbm 1} \right)\right),\;
      {\mathcal X}\in {\mathcal A} \otimes {\mathcal A},\; a\in {\mathcal A}
      \;.
    \end{equation}
    \item ${\mathcal E}:{\mathcal A}\rightarrow {\mathcal A}$ leaves the subspace ${\sf B}({\mathcal H})$ of Hermitean operators invariant
    iff  $C({\mathcal E}) \in {\sf B}({\mathcal H}\otimes {\mathcal H} )$.
    \item ${\mathcal E}:{\mathcal A}\rightarrow {\mathcal A}$ is trace-preserving iff $\mbox{Tr}_2 \left( C({\mathcal E})\right)={\mathbbm 1}$.
    \item ${\mathcal E}:{\mathcal A}\rightarrow {\mathcal A}$ is completely positive iff $C({\mathcal E}) \in {\sf B}^+({\mathcal H}\otimes {\mathcal H} )$.
  \end{enumerate}
\end{theorem}
For the proof see, e.~g., \cite{NC00} and \cite{FC24}. Note that the Choi isomorphism $C$ as well as its inverse $C^{-1}$
generally depend on the choice of an orthonormal basis in ${\mathcal H}$.
For $C$ this is obvious, for $C^{-1}$ it follows from the occurrence of
$a^\top$ in its definition (\ref{Cinverse}) and the transposition of an operator being base-dependent.

The orthonormal basis $\left(|i\rangle\right)_{0\le i<N}$ in ${\mathcal H}$ yields the product basis
$\left(|i k \rangle\right)_{0\le i,k<N}:=\left(|i\rangle\otimes |k\rangle\right)_{0\le i,k<N}$ in ${\mathcal H}\otimes {\mathcal H}$
and, further, the orthonormal basis $\left(|i k\rangle\langle j \ell|\right)_{0\le i,k,j,\ell<N}$ in $L({\mathcal H}\otimes {\mathcal H})$.
Expanding $C({\mathcal E})$ for any ${\mathcal E}\in L({\mathcal A})$ into the latter basis gives
\begin{equation}\label{expandCE}
 C({\mathcal E})= \sum_{0\le i k  j \ell<N}c^{i k,j \ell}\, |i k\rangle\langle j \ell|
 \;,
\end{equation}
thus defining the ``Choi matrix" $c({\mathcal E})\in L({\mathbb C}^{N\times N})$ with entries $c^{i k,j \ell}$.

\noindent In order to calculate $c({\mathcal E})$ we expand ${\mathcal E}(|i\rangle \langle j|)$
into the orthonormal basis $\left(|k\rangle \langle \ell|\right)_{0\le k,\ell<N}$
of ${\mathcal A}$:
\begin{equation}\label{expand Eij}
 {\mathcal E}(|i\rangle \langle j|)=\sum_{k \ell} {\mathcal E}_{ij}^{k\ell}\,|k\rangle \langle \ell|
 \;,
\end{equation}
and insert this expression into
\begin{eqnarray}
\label{CM1}
 C( {\mathcal E}) &\stackrel{(\ref{defChoi})}{=}&\left({\mathbbm 1}\otimes {\mathcal E} \right)\left( | \Phi\rangle\, \langle \Phi|  \right) \\
   &\stackrel{(\ref{defent})}{=}&\left({\mathbbm 1}\otimes {\mathcal E} \right)\left( \sum_{ij}\left(|i\rangle\langle j| \right) \otimes\left(|i\rangle\langle j| \right)\right)  \\
   \label{CM2}
   &=& \sum_{ij} \left( |i\rangle\langle j|\right) \otimes {\mathcal E}\left(  |i\rangle\langle j|\right)\\
   \label{CM3}
   &\stackrel{(\ref{expand Eij})}{=}&\sum_{ij} \left( |i\rangle\langle j|\right) \otimes \sum_{k\ell} {\mathcal E}_{ij}^{k\ell}  |k\rangle\langle \ell |\\
   \label{CM4}
   &=&\sum_{ijk\ell}  {\mathcal E}_{ij}^{k\ell}  |ik\rangle \langle j\ell|
   \;.
\end{eqnarray}
Comparison with (\ref{expandCE}) gives the result
\begin{equation}\label{CE}
  c^{i k,j \ell} = {\mathcal E}_{ij}^{k\ell}
  \;.
\end{equation}
Note that the matrix entries of $c$ and ${\mathcal E}$ are identical
except for the exchange of the indices $k \leftrightarrow j$.
On the matrix level the Choi isomorphism can therefore be understood as a sort of ``partial transposition",
see also Figure \ref{FIGCE} in Section \ref{sec:COI}.

Since the definition of $C({\mathcal E})$ presupposes an orthonormal basis in ${\mathcal H}$ anyway,
one can, without further loss of generality, also consider a ``Choi matrix isomorphism"
\begin{equation}\label{Cmatrixiso}
 c:  L({\mathcal A}) \rightarrow L({\mathbb C}^{N\times N})
\end{equation}
instead of the previously defined Choi isomorphism (\ref{defChoi}).
The properties of $c$ analogous to those of $C$, see Theorem \ref{T1},
are obvious and need not be repeated.

\section{The GKS isomorphism}\label{sec:GKSI}
\subsection{Choi isomorphism as a special case of the GKS isomorphism}\label{sec:GKSCI}

The GKS-isomorphism, named after the authors of \cite{GKS76}, was implicitly introduced in \cite{GKS76} in connection with
the time evolution of open quantum systems. For the convenience of the reader we will
reproduce the pertaining proofs that have been already published there.
Let $\left(F_\alpha \right)_{\alpha=0,\ldots,N^2-1}$ be an arbitrary orthonormal basis
in ${\mathcal A}$ w.~r.~t.~the scalar product (\ref{defscalarproductA}). Then one considers the family of linear maps
${\mathcal F}_{\alpha\beta}\in L({\mathcal A})$ defined by
\begin{equation}\label{defFab}
 {\mathcal F}_{\alpha\beta}(X) = F_\alpha\,X\,F_\beta^\ast,\quad X\in{\mathcal A},\quad 0\le \alpha,\beta< N^2
 \;.
\end{equation}
It is shown in \cite{GKS76} that this family actually defines a basis of $L({\mathcal A})$ but we will not use this fact
at the moment and postpone its proof.
Let us first focus on the special case that an orthonormal basis
$\left(|i\rangle\right)_{0\le i < N}$ of ${\mathcal H}$ is chosen and
consider the derived orthonormal basis $\left(|j\rangle\,\langle i|\right)_{0 \le i,j<N}$ of ${\mathcal A}$.
We write $\alpha\equiv(i,j)$ as a double index and set $ E_\alpha:= |j\rangle\,\langle i|$, hence
\begin{equation}\label{defEalpha}
  E_\alpha^\ast:= |i\rangle\,\langle j| \quad \mbox{for } 0\le i,j < N
  \;.
\end{equation}
Note the reversed positions of the indices.
Then it can be shown that
\begin{lemma}\label{L1}
The family  $\left({\mathcal E}_{\alpha\beta}\right)_{0\le \alpha,\beta<N^2}$
defined analogously to (\ref{defFab}) using the orthonormal basis $\left( E_\alpha\right)_{0\le \alpha< N^2}$ in ${\mathcal A}$
will be an orthonormal basis in $L({\mathcal A})$ w.~r.~t.~the scalar product (\ref{defscalarproductLA}).
\end{lemma}
For the {\bf proof} we will check that $\left({\mathcal E}_{\alpha\beta}\right)_{0\le \alpha,\beta<N^2}$ is orthonormal.
Then it must be a basis since its length is $N^4$, the dimension of the space $L({\mathcal A})$.
We use double indices
$\alpha=(i,j),\, \beta=(k,l),\,\gamma=(m,n),\,\delta=(p,q),\epsilon=(r,s)$ and consider:
\begin{eqnarray}
\label{ON1}
 \left\langle {\mathcal E}_{\alpha\beta}\left|\right. {\mathcal E}_{\gamma\delta}\right\rangle &=&
 \mbox{Tr } \left( {\mathcal E}_{\alpha\beta}^\ast\, {\mathcal E}_{\gamma\delta}\right) \\
  \label{ON2}
   &=& \sum_\epsilon \left\langle E_\epsilon \left|\right.
   {\mathcal E}_{\alpha\beta}^\ast\, {\mathcal E}_{\gamma\delta}\left|\right. E_\epsilon\right\rangle\\
  \label{ON3}
   &=& \sum_\epsilon \left\langle  {\mathcal E}_{\alpha\beta}\left( E_\epsilon\right) \left|\right.
   {\mathcal E}_{\gamma\delta}\left( E_\epsilon\right)\right\rangle\\
   \label{ON4}
   &=& \sum_\epsilon \left\langle E_\alpha\, E_\epsilon\,E_\beta^\ast \left|\right.
    E_\gamma\, E_\epsilon\,E_\delta^\ast\right\rangle               \\
   \label{ON5}
   &=&
    \sum_\epsilon \mbox{Tr }\left( E_\beta\, E_\epsilon^\ast\,E_\alpha^\ast\,
    E_\gamma\, E_\epsilon\,E_\delta^\ast\right)        \\
   \label{ON6}
   &=& \sum_{r,s}\mbox{Tr }\left(
   |l\rangle\langle k| r\rangle \langle s | i\rangle \langle j| n\rangle\langle m| s\rangle\langle r| p\rangle \langle q|
   \right)   \\
   \label{ON7}
   &=& \sum_{r,s} \delta_{lq}\,\delta_{kr}\,\delta_{si}\,\delta_{jn}\,\delta_{ms}\,\delta_{rp}      \\
   \label{ON8}
   &=&\delta_{lq}\,\delta_{kp}\,\delta_{im}\,\delta_{jn} = \delta_{\alpha\gamma}\,\delta_{\beta\delta}
   \;.
\end{eqnarray}
Hence $\left({\mathcal E}_{\alpha\beta}\right)_{0\le \alpha,\beta<N^2}$ is an orthonormal basis in $L({\mathcal A})$.
\hfill$\Box$\\

Let ${\mathcal E}\in L({\mathcal A})$. According to Lemma \ref{L1} it can be uniquely expanded w.~r.~t.~the basis
$\left({\mathcal E}_{\alpha\beta}\right)_{0\le \alpha,\beta<N^2}$, and hence there exist complex numbers $\widetilde{c}^{\alpha\beta}$
such that
\begin{equation}\label{expandE}
  {\mathcal E}(A) = \sum_{\alpha\beta} \widetilde{c}^{\alpha\beta} E_\alpha\,A\,E_\beta^\ast,\quad \mbox{for all } A\in {\mathcal A}
  \;.
\end{equation}
Then the matrix with entries $\widetilde{c}^{\alpha\beta}$ is exactly the Choi matrix of ${\mathcal E}$. More precisely, we have
\begin{prop}\label{P1}
 For any ${\mathcal E}\in L({\mathcal A})$ let the $N^2\times N^2$-matrix  $\widetilde{c}$ with entries  $\widetilde{c}^{\alpha\beta}$ be defined
 by (\ref{expandE}). Then
 \begin{equation}\label{ChoiGKS}
  \widetilde{c}= c({\mathcal E})
   \;.
 \end{equation}
\end{prop}
{\bf Proof}: We use double indices $\mu=(m,k)$ and $\nu=(n,\ell)$ and set $A=|i\rangle \langle j|$.
Then we obtain
\begin{eqnarray}
\label{proofa}
 {\mathcal E}\left(A\right)&=&
  {\mathcal E}\left( |i\rangle \langle j|\right) \stackrel{(\ref{expand Eij})}{=} \sum_{k \ell}  {\mathcal E}_{ij}^{k\ell}\,|k\rangle \langle \ell| \\
  \label{proofb}
   &\stackrel{(\ref{expandE})}{=}& \sum_{\mu\nu} \widetilde{c}^{\mu\nu}\,E_\mu\,A\,E_\nu^\ast\\
   \label{proofc}
   &=& \sum_{mkn\ell}\widetilde{c}^{mk,n\ell} \, |k\rangle \underbrace{\langle m| i\rangle}_{\delta_{mi}}
    \underbrace{\langle j| n \rangle}_{\delta_{jn}}\langle \ell|\\
    \label{proofd}
    &=& \sum_{k\ell} \widetilde{c}^{ik,j\ell} \, |k\rangle \langle \ell|
    \;.
\end{eqnarray}
Comparing (\ref{proofd}) and (\ref{proofa}) yields
\begin{equation}\label{proofe}
   \widetilde{c}^{ik,j\ell}= {\mathcal E}_{ij}^{k\ell}\stackrel{(\ref{CE})}{=}c^{ik,j\ell}
   \;,
\end{equation}
which completes the proof of (\ref{ChoiGKS}).   \hfill$\Box$\\

\subsection{Definition of the GKS isomorphisms}\label{sec:DGKSI}

In order to define the GKS-isomorphism we first will show that the family
$\left({\mathcal F}_{\alpha\beta}\right)_{0\le \alpha,\beta<N^2}$
defined in (\ref{defFab}) is always an orthonormal basis in $L({\mathcal A})$,
not only in the special case of $\left({\mathcal E}_{\alpha\beta}\right)_{0\le \alpha,\beta<N^2}$.
To this end we formulate the following
\begin{lemma}\label{L2}
Let $\left(F_\alpha\right)_{0\le \alpha<N^2}$ and $\left(G_\alpha\right)_{0\le \alpha<N^2}$ be two
orthonormal bases in ${\mathcal A}$ and $\left({\mathcal F}_{\alpha\beta}\right)_{0\le \alpha,\beta<N^2}$
and $\left({\mathcal G}_{\alpha\beta}\right)_{0\le \alpha,\beta<N^2}$ the corresponding families of superoperators
defined in (\ref{defFab}). Then the following holds:
  \begin{enumerate}
    \item There exists a unique unitary matrix $U\in {\sf U}({\mathbb C}^{N\times N})$ such that
    \begin{equation}\label{FUG}
      F_\alpha= \sum_{\alpha'}U_{\alpha\alpha'} G_{\alpha'}
      \;.
    \end{equation}
    \item The family $\left({\mathcal F}_{\alpha\beta}\right)_{0\le \alpha,\beta<N^2}$ is an orthonormal basis
    in $L({\mathcal A})$  iff $\left({\mathcal G}_{\alpha\beta}\right)_{0\le \alpha,\beta<N^2}$ is such a basis..
  \end{enumerate}
\end{lemma}
{\bf Proof:}
\begin{enumerate}
  \item Eq.~(\ref{FUG}) holds for complex numbers $U_{\alpha\alpha'}$ since $\left(G_\alpha\right)_{0\le \alpha<N^2}$
  is a basis in ${\mathcal A}$. Using the orthonormality of both bases we conclude
  \begin{eqnarray}
  \label{P21a}
    \delta_{\alpha\beta} &=& \left\langle F_\alpha\left|\right. F_\beta\right\rangle=
    \sum_{\alpha'\,\beta'}\overline{U_{\alpha\alpha'}}\,U_{\beta\beta'}\,\left\langle G_{\alpha'}\left|\right.G_{\beta'}\right\rangle
     \\
    \label{P21b}
     &=&  \sum_{\alpha'\,\beta'}\overline{U_{\alpha\alpha'}}\,U_{\beta\beta'}\,\delta_{\alpha'\,\beta'}=
     \sum_{\alpha'}\overline{U_{\alpha\alpha'}}\,U_{\beta\alpha'}
    =\sum_{\alpha'}U_{\beta\alpha'}\,U^\ast_{\alpha'\,\alpha}
    \;.
  \end{eqnarray}
  Hence $U\,U^\ast={\mathbbm 1}$ and $U$ is a unitary matrix.
  \item Both families of superoperators have the length $N^2$ which equals the dimension of the space $L({\mathcal A})$.
  We will assume that $\left({\mathcal G}_{\alpha\beta}\right)_{0\le \alpha,\beta<N^2}$ is orthonormal and show that then
  also $\left({\mathcal F}_{\alpha\beta}\right)_{0\le \alpha,\beta<N^2}$ is so. This would prove the second claim of the lemma.
  Hence we consider
  \begin{eqnarray}
  \label{P22a}
   \left\langle {\mathcal F}_{\alpha\beta}\left|\right. {\mathcal F}_{\gamma\delta}\right\rangle &=&
   \sum_\epsilon \mbox{Tr }\left( F_\beta\,F_\epsilon^\ast\,F_\alpha^\ast\, F_\gamma\,F_\epsilon\,F_\delta^\ast\right)
   \\
   \label{P22b}
    &=&  \sum_{\epsilon}\sum_{\alpha'\beta'\gamma'\delta'} \overline{U_{\alpha\alpha'}}\,U_{\beta\beta'}\,U_{\gamma\gamma'}\, \overline{U_{\delta\delta'}}\,
    \mbox{Tr }\left( G_{\beta'}\,F_\epsilon^\ast\,G_{\alpha'}^\ast\, G_{\gamma'}\,F_\epsilon\,G_{\delta'}^\ast\right)\\
   \label{P22c}
    &=& \sum_{\alpha'\beta'\gamma'\delta'} \overline{U_{\alpha\alpha'}}\,U_{\beta\beta'}\,U_{\gamma\gamma'}\, \overline{U_{\delta\delta'}}\,
    \underbrace{\left\langle {\mathcal G}_{\alpha'\beta'} \left|\right.{\mathcal G}_{\gamma'\delta'} \right\rangle}
    _{\delta_{\alpha'\gamma'}\delta_{\beta'\delta'}}\\
   \label{P22d}
    &=&\sum_{\alpha'\beta'}  \overline{U_{\alpha\alpha'}}\,U_{\beta\beta'}\,U_{\gamma\alpha'}\, \overline{U_{\delta\beta'}}=
    \sum_{\alpha'\beta'} U_{\gamma\alpha'}\, U_{\alpha'\alpha}^\ast\,U_{\beta\beta'}\, U_{\beta'\delta}^\ast =
    \delta_{\alpha\gamma}\,\delta_{\beta\delta}
    \;.
   \end{eqnarray}
 Here we have used in line (\ref{P22a}) an identity analogous to that leading from (\ref{ON1}) to (\ref{ON5}),
 and reversely in line  (\ref{P22c}).
\end{enumerate}
 \hfill$\Box$\\
From Lemma \ref{L2} 2.~and Lemma \ref{L1} we immediately conclude:
\begin{prop}\label{P2}
The family of superoperators
$\left({\mathcal F}_{\alpha\beta}\right)_{0\le \alpha,\beta<N^2}$
defined in (\ref{defFab}) is always an orthonormal basis in $L({\mathcal A})$.
\end{prop}

Now consider an arbitrary ${\mathcal E}\in L({\mathcal A})$. It can be expanded into the orthonormal basis
$\left({\mathcal F}_{\alpha\beta}\right)_{0\le \alpha,\beta<N^2}$ which yields

\begin{equation}\label{defGKSI}
  {\mathcal E}(A) = \sum_{0\le\alpha,\beta<N^2}g^{\alpha\beta}\, F_\alpha\,A\,F_\beta^\ast,\quad \mbox{for all } A\in{\mathcal A}
  \;,
\end{equation}
for some uniquely determined complex numbers $g^{\alpha\beta}$ that can be viewed as the entries of a matrix $g\in L({\mathbb C}^{N\times N})$.
This matrix will be denoted as the ``GKS matrix" and used to define the GKS-isomorphism
\begin{eqnarray}
\label{defGKSIa}
  G &:& L({\mathcal A}) \rightarrow L({\mathbb C}^{N\times N}),\\
  \label{defGKSIb}
   && {\mathcal E} \mapsto G({\mathcal E}):=g
   \;.
\end{eqnarray}

Conversely, every matrix $g\in L({\mathbb C}^{N\times N})$ generates a superoperator ${\mathcal E}\in  L({\mathcal A}) $
by means of (\ref{defGKSI}). Hence it is clear that
$ G: L({\mathcal A}) \rightarrow L({\mathbb C}^{N\times N})$ is a linear isomorphism.

The above definition of the GKS-isomorphism is an implicit one. To obtain an explicit form,
let us describe the superoperator ${\mathcal E}\in  L({\mathcal A})$ by means of an $N^2\times N^2$-matrix $M$
in the usual way:
\begin{equation}\label{defM}
  {\mathcal E}(F_\alpha) = \sum_\beta M_{\beta \alpha} F_\beta,\quad \mbox{for } 0\le  \alpha< N^2
  \;.
\end{equation}
Further define a rank $4$ tensor $\Gamma$ by
\begin{equation}\label{defGamma}
\Gamma^{\delta \gamma \sigma \tau} := \mbox{Tr} \left(F_\delta F_\sigma^\ast F_\gamma^\ast F_\tau \right), 
\quad \mbox{for } 0\le \delta,\sigma,\gamma,\tau< N^2
\;.
\end{equation}
Then, after some calculations, the implicit definition of the GKS-isomorphism can be transformed into the following
\begin{defi}\label{DGKS}
 Let $\left(F_\alpha \right)_{\alpha=0,\ldots,N^2-1}$ denote an orthonormal basis in ${\mathcal A}$ w.~r.~t.~the 
 scalar product (\ref{defscalarproductA}).
 Then the GKS-matrix isomorphism $M\mapsto g$ is defined by the equation
 \begin{equation}\label{defGKSMI}
  g^{\delta \gamma} := \sum_{\sigma\tau} \Gamma^{\delta \gamma \sigma \tau} \; M_{\sigma\tau},\quad \mbox{for } 0\le \delta,\gamma< N^2
  \;,
  \end{equation}
  with $\Gamma$ defined by (\ref{defGamma}) and $M$ by (\ref{defM}).
\end{defi}

Sometimes the term ``GKS matrix" appears in the literature with a slightly different meaning, 
namely in connection with the representation of the Lindblad generators of dynamic semigroups, 
see for example \cite{Betal01}. We will provide a note on this in the appropriate place, in subsection \ref{sec:TDLE}.

The comparison of (\ref{defGKSI}) and (\ref{expandE}) shows that the Choi matrix isomorphism is a special
case of the GKS-isomorphism. Already the Choi matrix isomorphism is not a single mathematical object but rather a family
since it depends on a family of orthonormal bases in ${\mathcal A}$ that are of the form $\left(|i\rangle \langle j|\right)_{0\le i,j<N}$
w.~r.~t.~the family of orthonormal bases in ${\mathcal H}$ denoted by $\left( |i\rangle\right)_{0\le i <N}$.
The GKS-isomorphism $G$ further extends this family by considering general orthonormal bases
$\left(F_\alpha\right)_{0\le \alpha <N^2}$ in ${\mathcal A}$.
Of course, it still has to be shown that $G$  has analogous
properties as the Choi matrix isomorphism.

\subsection{Properties of the GKS isomorphisms}\label{sec:PGKSI}
First we investigate the transformation of the GKS matrix $g=G({\mathcal E})$
under changes of the orthonormal basis $F_\alpha \mapsto G_\alpha$ of ${\mathcal A}$
given by  a unitary matrix $U$ according to (\ref{FUG}). Let $V=U^\top$ denote
the transpose of $U$, again a unitary matrix satisfying $V^\ast=\overline{U}$.
Let $A\in{\mathcal A}$ then we conclude
\begin{eqnarray}
\label{transGKSmatrixa}
  {\mathcal E}(A)&=&\sum_{\alpha\beta}g^{\alpha\beta}\, F_\alpha\,A\,F_\beta^\ast \\
  \label{transGKSmatrixb}
   &=& \sum_{\alpha\alpha'\beta\beta'}g^{\alpha\beta}\, U_{\alpha\alpha'}\, G_{\alpha'}\,A\,G_{\beta'}^\ast\,\overline{U_{\beta\beta'}}
   \\
  \label{transGKSmatrixc}
   &=&\sum_{\alpha'\beta'}\left( \sum_{\alpha\beta} V_{\alpha'\alpha}\,g^{\alpha\beta}\,V_{\beta\beta'}^\ast
   \right) \, G_{\alpha'}\,A\,G_{\beta'}^\ast \\
  \label{transGKSmatrixd}
   &=:& \sum_{\alpha'\beta'} \hat{g}^{\alpha'\beta'}\,G_{\alpha'}\,A\,G_{\beta'}^\ast
   \;.
\end{eqnarray}
Hence the GKS matrix transforms according to $g\mapsto \hat{g}= V\,g\,V^\ast$.
This implies that any property of the GKS matrix that is invariant under unitary transformations
holds for all orthonormal bases $\left(F_\alpha\right)_{0\le \alpha <N^2}$ of ${\mathcal A}$ if it holds
for a particular one. In this way the GKS matrix isomorphism  $G : L({\mathcal A}) \rightarrow L({\mathbb C}^{N\times N})$
inherits the corresponding properties of the Choi matrix isomorphism.
This entails
\begin{theorem}\label{T2}
Let ${\mathcal E}\in L({\mathcal A})$ and  $g=G({\mathcal E})$. Then the following holds:
  \begin{enumerate}
  \item ${\mathcal E}:{\mathcal A}\rightarrow {\mathcal A}$ leaves the subspace ${\sf B}({\mathcal H})$ of Hermitean operators invariant
    iff  $G({\mathcal E}) \in {\sf B}({\mathbb C}^{N\times N})$.
  \item ${\mathcal E}:{\mathcal A}\rightarrow {\mathcal A}$ is trace-preserving iff
  $\sum_{\alpha\beta}\overline{g^{\alpha\beta}}\,F_\alpha^\ast\,F_\beta={\mathbbm 1}$.
  \item ${\mathcal E}:{\mathcal A}\rightarrow {\mathcal A}$ is completely positive iff $G({\mathcal E}) \in {\sf B}^+({\mathbb C}^{N\times N})$.
  \end{enumerate}
\end{theorem}
It will be instructive to check those parts of this Theorem which differ in the formulation from Theorem \ref{T1}.
We will only consider the case where ${\mathcal E}$
leaves the subspace ${\sf B}({\mathcal H})$ of Hermitean operators invariant and is trace-preserving.
Then we conclude that $g=G({\mathcal E})$ is Hermitean, i.~e., $\overline{g^{\alpha\beta}}=g^{\beta\alpha}$ and further
\begin{eqnarray}
\label{PT21a}
  \mbox{Tr } A &=&  \mbox{Tr } {\mathcal E}(A)= \sum_{\alpha\beta} g^{\alpha\beta}\, \mbox{Tr } \left(F_\alpha\,A\,F_\beta^\ast \right)
  \quad\mbox{for all } A\in{\mathcal A}\\
  \label{PT21b}
   \Leftrightarrow && \mbox{Tr } \left(\left(\sum_{\alpha\beta} g^{\alpha\beta}\,F_\beta^\ast\,F_\alpha\right)-{\mathbbm 1}\right) A
   =0 \quad\mbox{for all } A\in{\mathcal A}\\
   \label{PT21c}
    \Leftrightarrow && \sum_{\alpha\beta} g^{\alpha\beta}\,F_\beta^\ast\,F_\alpha={\mathbbm 1}
   \Leftrightarrow \sum_{\beta\alpha} g^{\beta\alpha}\,F_\alpha^\ast\,F_\beta={\mathbbm 1}
   \Leftrightarrow \sum_{\beta\alpha} \overline{g^{\alpha\beta}}\,F_\alpha^\ast\,F_\beta={\mathbbm 1}
   \;,
\end{eqnarray}
which confirms Theorem \ref{T2}.2.

Concerning Theorem \ref{T2}.3. we consider the case where $g=G({\mathcal E})\ge 0$ and adopt its spectral decomposition in the form
\begin{equation}\label{spectralg}
 g= W\,\Delta\,W^\ast
 \;,
\end{equation}
such that $\Delta$ is a diagonal matrix containing the eigenvalues $g^\alpha\ge 0$ of $g$
and $W$ is a unitary matrix the columns of which are the corresponding eigenvectors.
It follows that for all $X\in{\mathcal A}$
\begin{eqnarray}
\label{com1}
  {\mathcal E}(X)&=&\sum_{\alpha\beta}g^{\alpha\beta}\, F_\alpha\,X\,F_\beta^\ast \\
   &=& \sum_{\alpha\beta\gamma\lambda} W_{\alpha\gamma}\,\underbrace{\Delta_{\gamma\lambda}}_{g^\gamma \delta_{\gamma\lambda}}
   \,W_{\lambda\beta}^\ast \, F_\alpha\,X\,F_\beta^\ast \\
   &=& \sum_{\alpha\beta\gamma} W_{\alpha\gamma}\, g^\gamma  \,W_{\gamma\beta}^\ast \, F_\alpha\,X\,F_\beta^\ast \\
   &=& \sum_\gamma g^\gamma \,\left(\sum_\alpha W_{\alpha\gamma}F_\alpha\right) \,X\,\left(\sum_\beta \overline{W_{\beta\gamma}}F_\beta^\ast \right) \\
   &=:& \sum_\gamma g^\gamma \, A_\gamma\,X\,A_\gamma^\ast   \;.
\end{eqnarray}
This gives a representation of ${\mathcal E}(X)$ in terms of the Kraus operators
\begin{equation}\label{Kraus}
 \widetilde{A}_\gamma := \sqrt{g^\gamma}\,A_\gamma:= \sqrt{g^\gamma}\,\sum_\alpha W_{\alpha\gamma}F_\alpha
 \;,
\end{equation}
thereby again proving that  ${\mathcal E}$ is completely positive.

As a simple example we consider the  orthonormal (Pauli) basis
\begin{equation}\label{GKSbasisN2}
 F_0=\frac{1}{\sqrt{2}}\left( \begin{array}{cc}
              1& 0 \\
              0 & 1
            \end{array}\right),\;
 F_1=\frac{1}{\sqrt{2}}\left( \begin{array}{cc}
             0& 1 \\
             1 & 0
            \end{array}\right),\;
 F_2=\frac{1}{\sqrt{2}}\left( \begin{array}{cc}
              0& -{\sf i} \\
              {\sf i} & 0
            \end{array}\right),\;
 F_3=\frac{1}{\sqrt{2}}\left( \begin{array}{cc}
              1& 0 \\
              0 & -1
            \end{array}\right),
\end{equation}
in ${\mathcal A}$ and the corresponding single qubit ``Pauli channel"
\begin{equation}\label{Pauli channel}
{\mathcal E}(\rho) = \sum_{\alpha=0}^{3} k_\alpha F_\alpha \,\rho F_\alpha
\;,
\end{equation}
where $k_\alpha>0$ for $\alpha=0,1,2,3$. It is trace-preserving iff $\sum_{\alpha=0}^{3}k_\alpha=2$.
Obviously, the GKS-matrix of ${\mathcal E}$ is the diagonal matrix with entries $k_0,k_1,k_2,k_3$,
which is positive in accordance with the fact that the Pauli channel is completely positive.

\section{Comparison with other isomorphisms}\label{sec:COI}

\begin{figure}[h]
  \centering
    \includegraphics[width=0.5\linewidth]{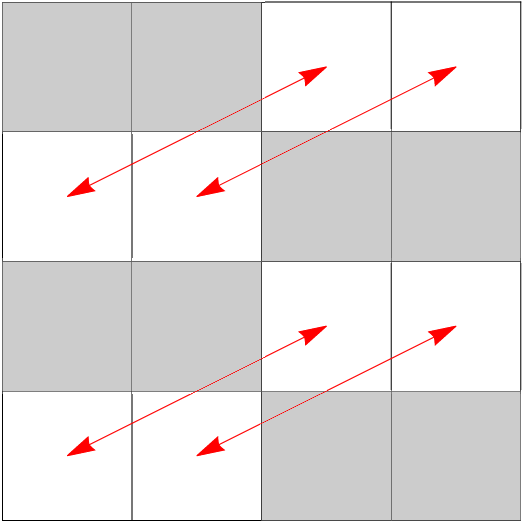}
  \caption[CE]
  {Illustration of the Choi matrix isomorphism $c({\mathcal E})$ for the case of $N=2$.
  It performs a permutation of the matrix elements of the $4\times 4$ matrix ${\mathcal E}$,
  which consists of four transpositions marked by red double arrows.  The matrix elements in the gray squares remain fixed.
  }
  \label{FIGCE}
\end{figure}

There exist various other generalizations of the Choi isomorphism and the question arises whether the
GKS-isomorphism considered above is really different. We will check this by an application to the example
of the transposition $\tau$  of $2\times 2$-matrices considered in Section \ref{sec:Def} which maps positive matrices to
positive ones but is not completely positive.

First, we will consider the application of the Choi matrix isomorphism and the GKS isomorphism to this example.
Let  $N=2$ and consider the orthonormal basis
\begin{equation}\label{basisN2}
 E_0=\left( \begin{array}{cc}
              1& 0 \\
              0 & 0
            \end{array}\right),\;
             E_1=\left( \begin{array}{cc}
             0& 1 \\
              0 & 0
            \end{array}\right),\;
             E_2=\left( \begin{array}{cc}
              0& 0 \\
              1 & 0
            \end{array}\right),\;
             E_3=\left( \begin{array}{cc}
              0& 0 \\
              0 & 1
            \end{array}\right),
\end{equation}
in ${\mathcal A}$. Since $\tau$ only swaps $E_1$ and $E_2$ we can represent it by the matrix
\begin{equation}\label{matrixphi}
 \widetilde{\tau} =\left(
 \begin{array}{cccc}
   1 & 0 & 0 & 0 \\
   0 & 0 & 1 & 0 \\
   0 & 1 & 0 & 0 \\
   0 & 0 & 0 & 1
 \end{array}
 \right)
 \;.
\end{equation}
According to (\ref{CE}) the Choi matrix $c(\tau)$ of ${\tau}$ is obtained by a certain permutation
of matrix elements, see Figure \ref{FIGCE}. It follows that $c(\tau)=\widetilde{\tau}$ and hence $c({\tau})$
is not positively semi-definite in accordance with the fact that $\tau$ is not completely positive.

We will perform the same check with the GKS isomorphism corresponding to the orthonormal (Pauli) basis
(\ref{GKSbasisN2}) in ${\mathcal A}$. After some calculations we obtain the following  GKS-matrix $G(\tau)$
\begin{equation}\label{GKSphi}
 G(\tau)=\left(
\begin{array}{cccc}
 1 & 0 & 0 & 0 \\
 0 & 1 & 0 & 0 \\
 0 & 0 & -1 & 0 \\
 0 & 0 & 0 & 1 \\
\end{array}
\right)
\;,
\end{equation}
which is not positively semi-definite either.

\subsection{Comparison with dePillis-Jamio{\l}ski isomorphism}\label{sec:COG1}

As already noted in \cite{PS13} and again in \cite{FC24}, the isomorphism due to \cite{dP67} and \cite{J72}
differs from the Choi isomorphism, although this difference is sometimes
blurred by the misleading denotation ``Choi-Jamio{\l}ski isomorphism".
It can be written in the form
\begin{equation}\label{PL}
  J({\mathcal E})=\sum_\alpha F_\alpha^\ast\otimes{\mathcal E}(F_\alpha)
  \;,
\end{equation}
seemingly depending on an orthonormal basis $\left(F_\alpha\right)_{0\le \alpha <N^2}$ in ${\mathcal A}$, but  $J({\mathcal E})$
can be shown to be independent of this basis, see \cite{FC24}.
Again we calculate the matrix of $J(\tau)$ where $\tau$ denotes the transposition of $2\time 2$-matrices
and obtain
\begin{equation}\label{dPJphi}
 J(\tau)=\left(
\begin{array}{cccc}
 1 & 0 & 0 &1 \\
 0 & 0 & 0 & 0 \\
 0 & 0 & 0 & 0 \\
1 & 0 & 0 & 1 \\
\end{array}
\right)
\;,
\end{equation}
which is positively semi-definite with eigenvalues $(2,0,0,0)$. Hence the dePillis-Jamio{\l}ski isomorphism
does not provide a criterion for complete positivity.

\subsection{Comparison with Paulsen-Shultz-Kye-Han isomorphism}\label{sec:COG2}
Recall that the Choi isomorphism can be written in the form
\begin{equation}\label{C1}
  C({\mathcal E})= \sum_{0\le i,j<N} |i\rangle\langle j | \otimes {\mathcal E}(|i\rangle\langle j | )
  \;,
\end{equation}
see (\ref{CM2}). Hence an obvious generalization consists in replacing the
orthonormal basis $\left(  |i\rangle\langle j |\right)_{0\le i,j<N}$ of ${\mathcal A}$
by a general one, say, $\left( F_\alpha\right)_{0\le \alpha<N^2}$. This idea has been
worked out in \cite{PS13}, therefore we will call the resulting isomorphism
\begin{equation}\label{PS}
 \mbox{PS}: L\left( L\left({\mathbbm C}^N \right),  L\left({\mathbbm C}^M \right)\right)
 \rightarrow \left({\mathbbm C}^N \right)\otimes \left({\mathbbm C}^M \right)
\end{equation}
the ``Paulsen-Shultz isomorphism" after the authors of this paper.

For $N=M=2$, the orthonormal basis $(F_0,F_1,F_2,F_3)$ defined in (\ref{GKSbasisN2})
and satisfying $F_\alpha^\ast =F_\alpha$,
and any linear map ${\mathcal E}:  L\left({\mathbbm C}^2\right) \rightarrow L\left({\mathbbm C}^2\right)$
we obtain for $\mbox{PS}({\mathcal E})$
the same result as for the  dePillis-Jamio{\l}ski isomorphism (\ref{dPJphi}).
The Paulsen-Shultz isomorphism hence provides a criterion for complete positivity
that applies to some orthonormal bases, but not to all. In fact, Paulsen and Shultz
provide a condition for a general basis such that the Paulsen-Shultz isomorphism
provides a criterion for complete positivity analogous to the the Choi criterion, see
corollary 12 in \cite{PS13}, first item.

A further generalization is achieved in \cite{K22} ,\cite{K22} and \cite{HK24} by considering the matrix
$C_\phi^\sigma$ defined by
\begin{equation}\label{defCphisigma}
C_\phi^\sigma = \sum_{ij} e_{i,j}\otimes \phi(\sigma(e_{i,j}))
\;,
\end{equation}
where $e_{i,j}=|i\rangle \langle j|$ denotes the standard matrix units
and $\sigma\in L({\mathcal A})$ is given by $\sigma(A) = s A s^\ast$
for some invertible operator $s\in{\mathcal A}$.
Here, too, the dependence on the special basis $e_{i,j}$ can be overcome
since $C_\phi^\sigma$ depends only on the bilinear form $\langle , \rangle$, which is given by
$\langle e_{i,j} ,\sigma(e_{k,l}) \rangle= \delta_{ik}\delta_{jl}$.
The linear transformation $\phi\mapsto C_\phi^\sigma$ will be called the
``Paulsen-Shultz-Kye-Han isomorphism".
It fulfills a refined and graded criterion for the k-positivity of $\phi$,
see theorem 4.2 in \cite{HK23}.

Next we will consider the relation to the GKS-isomorphism.
Let $U\in{\sf U}\left( {\mathbb C}^N\right)$ be a unitary $N\times N$ matrix and $g\in L({\mathbb C}^{N\times N})$
the GKS-matrix of $\phi\in L({\mathcal A})$ w.~r.~t.~the orthonormal basis $\left( E_\alpha U\right)_{0\le \alpha<{N\times N}}$ in ${\mathcal A}$
and $E_\alpha$ defined in (\ref{defEalpha}).
It follows that
\begin{eqnarray}
\label{GKSPS1}
  \phi(U^\ast A U) &=&\sum_{\alpha \beta} g^{\alpha\beta}\, \left( E_\alpha U\right)\, \left(U^\ast A U \right)\,\left( E_\beta U\right)^\ast\\
  \label{GKSPS1}
   &=& \sum_{\alpha \beta} g^{\alpha\beta}\,E_\alpha \,A\,E_\beta^\ast
   \;.
\end{eqnarray}
Hence the GKS-matrix $g$ coincides with the Choi matrix $C_\phi^\sigma$ in \cite{HK23}
for $\sigma(A)=U^\ast A U $.

Summarizing, we conclude that for the special choice of orthonormal bases of the form
 $\left( E_\alpha U\right)_{0\le \alpha<N}$ the GKS-isomorphism and the PSKH-isomorphism
 coincide, but in general they represent different generalizations of the Choi matrix isomorphism.

\subsection{Comparison with Frembs-Cavalcanti isomorphism}\label{sec:COG3}
In \cite{FC24} Frembs and Cavalcanti consider the Jordan product $\{a,b\}:=a\, b+b\,a$ and two different associative
products on  ${\mathcal A}$,
namely
\begin{eqnarray}
\label{productplus}
  a\cdot_+ b&=& \frac{1}{2}\{a,b\}+\frac{1}{2}\left[ a,b\right]= a\,b \\
  \label{productminus}
 a\cdot_- b&=& \frac{1}{2}\{a,b\}-\frac{1}{2}\left[ a,b\right]=b\,a
 \;.
\end{eqnarray}
The linear space ${\mathcal A}$ equipped with the first product (\ref{productplus}) can be identified
with the $C^\ast$-algebra ${\mathcal A}$, the second product (\ref{productminus}) defines a
different $C^\ast$-algebra denoted by  ${\mathcal A}^{\rm{op}}$.
For both $C^\ast$-algebras the ``star" is given by the adjoint map
$\ast: {\mathcal A}\rightarrow{\mathcal A}$ which maps each operator onto its adjoint.
There exist anti-linear isomorphisms
between both $C^\ast$-algebras, for example, the adjoint map.
With respect to a chosen orthonormal basis in ${\mathcal H}$  the transposition $\tau$
can be defined and represents a linear isomorphism  $\tau: {\mathcal A}^{\rm{op}} \rightarrow  {\mathcal A}$.

With these definitions Frembs and Cavalcanti prove the following
theorem (theorem  6 in \cite{FC24}):
\begin{theorem}\label{TFC}
 Let $\Phi^\ast\in L({\mathcal A})$ and $\left( F_\alpha\right)_{0\le \alpha<N^2}$ be an orthonormal basis
 in ${\mathcal A}$. Then
 \begin{equation}\label{rho}
  \rho:= \sum_\alpha F_\alpha^\ast \otimes \Phi^\ast\left(F_\alpha \right)\in {\mathcal A}^{\rm{op}}\otimes {\mathcal A}
 \end{equation}
 is positive iff $\Phi^\ast$ is completely positive. Moreover, $\rho$ does not depend on the orthonormal
 basis $\left( F_\alpha\right)_{0\le \alpha<N^2}$.
\end{theorem}
We have reformulated and adapted this theorem to the case considered in this paper.

In the following we will evaluate the condition that $\rho\in{\mathcal A}^{\rm{op}}\otimes {\mathcal A}$ is positive.
Recall that an element $X$ of a $C^\ast$-algebra ${\mathcal C}$ is defined as ``positive" , or $X\ge 0$, iff
$X$ is of the form $X=A\,A^\ast$ with $A\in {\mathcal C}$. A general element $A$ of the $C^\ast$-algebra
${\mathcal A}^{\rm{op}}\otimes {\mathcal A}$ can be written as a finite sum of the form
\begin{equation}\label{X1}
  A = \sum_\alpha c_\alpha \,a_\alpha\otimes b_\alpha,\quad \mbox{where } a_\alpha\in {\mathcal A}^{\rm{op}},\;
  b_\alpha\in {\mathcal A},\;\mbox{and } c_\alpha\in {\mathbb C}
  \;.
\end{equation}
Therefore an element $X\in {\mathcal A}^{\rm{op}}\otimes {\mathcal A}$ is positive iff it is of the form
\begin{eqnarray}
\label{positive1}
  X &=& A \,A^\ast  =\left(  \sum_\alpha c_\alpha\, a_\alpha\otimes b_\alpha  \right)
                     \left(  \sum_\beta \overline{c_\beta}\, a_\beta^\ast\otimes b_\beta^\ast  \right)  \\
  \label{positive2}
   &=& \sum_{\alpha\beta} c_\alpha\,\overline{c_\beta}\,  a_\beta^\ast\,a_\alpha \otimes b_\alpha\,b_\beta^\ast
   \;,
\end{eqnarray}
where we have used the two definitions (\ref{productplus}) and (\ref{productminus}) for the products
in ${\mathcal A}$ and ${\mathcal A}^{\rm{op}}$, resp.~. It is thus clear that this condition is, in general,
different from $X\ge 0$ for $X\in {\mathcal A}\otimes {\mathcal A}$. We consider the map
\begin{equation}\label{phi1}
 \tau\otimes {\mathbbm 1}:{\mathcal A}^{\rm{op}}\otimes {\mathcal A}\rightarrow {\mathcal A}\otimes {\mathcal A}
 \;,
\end{equation}
where $\tau$ is the linear isomorphism $\tau: {\mathcal A}^{\rm{op}} \rightarrow  {\mathcal A}$ given
by the transposition of matrices. It follows that for $X\ge 0$ where $X\in {\mathcal A}^{\rm{op}}\otimes {\mathcal A}$
we have
\begin{eqnarray}
  (\tau\otimes {\mathbbm 1})(X) &\stackrel{(\ref{positive2})}{=}&
  \sum_{\alpha\beta} c_\alpha\,\overline{c_\beta}\,  \tau(a_\beta^\ast\,a_\alpha) \otimes b_\alpha\,b_\beta^\ast \\
   &=&  \sum_{\alpha\beta} c_\alpha\,\overline{c_\beta}\,  a_\alpha^\top\,a_\beta^{\top\ast} \otimes b_\alpha\,b_\beta^\ast
   \;,
\end{eqnarray}
and hence $( \tau\otimes {\mathbbm 1})(X)\ge 0$ in $ {\mathcal A}\otimes  {\mathcal A}$. The converse can be shown analogously.

Therefore $\rho\in{\mathcal A}^{\rm{op}}\otimes {\mathcal A}$ is positive iff
$(\tau\otimes {\mathbbm 1})(\rho)\in{\mathcal A}\otimes {\mathcal A}$ is positive.
Inserting this result into Theorem \ref{TFC} and using that $\tau\left( F_\alpha^\ast\right)=\overline{ F_\alpha}$
we obtain the following:
\begin{theorem}\label{TFCM}
 Let $\Phi\in L({\mathcal A})$ and $\left( F_\alpha\right)_{0\le \alpha<N^2}$ be an orthonormal basis
 in ${\mathcal A}$. Then
 \begin{equation}\label{rhoM}
  \rho(\Phi):= \sum_\alpha \overline{F_\alpha} \otimes \Phi\left(F_\alpha \right)\in {\mathcal A}\otimes {\mathcal A}
 \end{equation}
 is positive iff $\Phi$ is completely positive.
 The complex conjugate matrix $\overline{F_\alpha}$ in (\ref{rhoM}) is defined w.~r.~t.~some fixed, arbitrary orthonormal basis
 $(|i\rangle)_{0\le i<N}$ in ${\mathcal H}$.
 Moreover, $\rho(\Phi)$ does not depend on the orthonormal
 basis $\left( F_\alpha\right)_{0\le \alpha<N^2}$.
\end{theorem}
For simplicity, we will also call the isomorphism $\Phi\mapsto \rho(\Phi)$ the  ``Frembs-Cavalcanti isomorphism"
and denote $\rho(\Phi)$ in (\ref{rhoM}) by $\mbox{FC}(\Phi)$.
It is essentially the dePillis-Jamio{\l}ski isomorphism followed by the block transposition $\tau\otimes {\mathbbm 1}$.
In the Theorem \ref{TFCM} we got rid of the reference to the somewhat unfamiliar $C^\ast$-algebra ${\mathcal A}^{\rm{op}}$,
but at the price of getting a matrix $\mbox{FC}(\Phi)$ instead of an operator.
The result of Theorem \ref{TFCM} is not new but can already be found in \cite{PS13}, theorem 22.
In this sense the Frembs-Cavalcanti isomorphism is equivalent to a variant of the Poulsen-Shultz-isomorphism.
For an earlier source, see also \cite{S01}, eq.~(8).

For the example $N=2$ and $\Phi=\tau$ (transposition of matrices) we obtain for both orthonormal bases
(\ref{basisN2}) and (\ref{GKSbasisN2}) the same matrix $\mbox{FC}(\tau)$ as for the Choi isomorphism (\ref{matrixphi}).
It is not positively semi-definite, in line with the fact that $\tau$ is not completely positive.
On the other hand, $\mbox{FC}(\tau)$ is different from the GKS-matrix $G(\tau)$ given by (\ref{GKSphi}),
which shows that both isomorphisms, $\mbox{FC}$ and $\mbox{GKS}$,  are different.

\section{Application to the time evolution of open quantum systems}\label{sec:A}

The time-independent GKSL equation for describing the dynamics of open quantum systems 
in the Markov approximation has numerous physical applications, which we cannot discuss here; 
for an overview, see, for example, \cite{AL07}. 
Instead, in this paper we will focus on the \textit{time-dependent} GKSL equation
and GKSL isomorphism.

\subsection{Time-dependent GKS isomorphism}\label{sec:TDGKS}

Motivated by the investigations of the dynamics of open quantum systems
we consider a time dependent, trace-preserving completely positive map ${\mathcal V}(t)$ that operates on
statistical operators $\rho_0\in {\sf B}^+_1({\mathcal H})$. The time-dependence is assumed to be sufficiently smooth.
According to Theorem \ref{T2} we conclude
\begin{equation}\label{evolution}
  \rho(t):= {\mathcal V}(t)\,\rho_0 = \sum_{\alpha\beta}g^{\alpha\beta}(t)\, F_\alpha\,\rho_0\,F_\beta^\ast
  \;,
\end{equation}
where $\left(F_\alpha\right)_{0\le \alpha<N^2}$ is a fixed orthonormal basis in ${\mathcal A}$,
$ g^{\alpha\beta}(t)$ are the entries of a positively semi-definite matrix $g(t)\in {\sf B}^+\left({\mathbb C}^{N\times N}\right)$,
and the time $t$ runs through some interval $0\le t < T$ including the possible case of $T=\infty$.
Differentiating (\ref{evolution}) w.~r.~t.~time $t$ gives
\begin{equation}\label{devolutiondt}
 \frac{d}{dt} \rho(t) = \sum_{\alpha\beta}\frac{d}{dt}{g}^{\alpha\beta}(t)\, F_\alpha\,\rho_0\,F_\beta^\ast
  \;.
\end{equation}
On the other hand, one can consider the linear superoperator
\begin{equation}\label{dV}
{\mathcal K}(t):= \left(\frac{d}{dt}{\mathcal V}(t)\right)\,{\mathcal V}^{-1}(t)
\;,
\end{equation}
assuming that ${\mathcal V}(t)$ is invertible for $0\le t<T$. ${\mathcal K}(t)$ maps
${\sf B}(\mathcal H)$ into ${\sf B}_0(\mathcal H)$ (thus being ``trace-annihilating"),
but cannot longer be assumed as completely positive. Again, Theorem \ref{T2} yields the
representation
\begin{equation}\label{Krep}
 \frac{d}{dt} \rho(t) = {\mathcal K}(t)  \rho(t) = \sum_{\alpha\beta}{\sf k}^{\alpha\beta}(t)\, F_\alpha\,\rho(t)\,F_\beta^\ast
 \;,
\end{equation}
where the ${\sf k}^{\alpha\beta}(t)$ are the entries of an Hermitean matrix ${\sf k}(t) \in{\sf B}\left({\mathbb C}^{N\times N}\right)$,
the GKS-matrix of ${\mathcal K}(t) $, which, however, cannot longer be assumed to have only non-negative eigenvalues.
We will refer to (\ref{Krep}) as the ``GKS equation".

The condition of ${\mathcal K}(t)$ being trace-annihilating can be evaluated further and gives
\begin{eqnarray}\label{traceannia}
0&= &\frac{d}{dt}\mbox{Tr } \rho(t)\stackrel{(\ref{Krep})}{=} \mbox{Tr }\left(\underbrace{ \left(
\sum_{\alpha\beta}{\sf k}^{\alpha\beta}(t) F_\beta^\ast\,F_\alpha\right)}_0
\rho(t)\right),\quad \mbox{for all } \rho(t)\;,\\
\label{traceannib}
&\Rightarrow& 0=\sum_{\alpha\beta}{\sf k}^{\alpha\beta}(t) F_\beta^\ast\,F_\alpha\;,\\
\label{traceanniac}
&\Rightarrow&0= \sum_{\alpha\beta}{\sf k}^{\alpha\beta}(t)\underbrace{\mbox{Tr }\left( F_\beta^\ast\,F_\alpha\right)}_{\delta_{\alpha\beta}}
=\mbox{Tr }{\sf k}(t)
\;.
\end{eqnarray}

Our goal is to transform (\ref{Krep}) to the level of GKS matrices.
The equations (\ref{devolutiondt}) and (\ref{Krep}) cannot be directly compared since  (\ref{devolutiondt}) expresses
$\frac{d}{dt} \rho(t)$ in terms of the ``initial value" $\rho_0$, whereas (\ref{Krep}) expresses it in terms of the
``actual value" $\rho(t)$. To overcome this problem we insert (\ref{evolution}) into (\ref{Krep}) with the result
\begin{equation}\label{Krevolution}
  \frac{d}{dt} \rho(t) =
   \sum_{\alpha\beta\gamma\delta} {\sf k}^{\gamma\delta}\,g^{\alpha\beta}\, F_\gamma\,F_\alpha\,\rho_0\,F_\beta^\ast\,F_\delta^\ast
  \;.
\end{equation}
To move on, we express the multiplication law in ${\mathcal A}$ in the orthonormal basis as
\begin{equation}\label{multFF}
 F_\gamma\,F_\alpha= \sum_\lambda \Pi ^\lambda_{\gamma\alpha} F_\lambda
 \;,
\end{equation}
with some time-independent array of complex numbers $\Pi ^\lambda_{\gamma\alpha}$.
This entails
\begin{equation}\label{multFastFast}
 F_\beta^\ast\,F_\delta^\ast= \sum_\mu \overline{\Pi ^\mu_{\delta\beta}} F_\mu
 \;,
\end{equation}
and, further,
\begin{equation}\label{KrevoPi}
  \frac{d}{dt} \rho(t) =\sum_{\lambda\mu}\underbrace{\left(
  {\sf k}^{\gamma\delta}\,g^{\alpha\beta}\, \Pi^\lambda_{\gamma\alpha}\,\overline{\Pi^\mu_{\delta\beta}}
  \right)
  }_{=:\hat{\sf k}^{\lambda\mu}}\,
  F_\lambda\,\rho_0\, F_\mu^\ast
  \;.
\end{equation}
Comparison with (\ref{devolutiondt}) now yields
\begin{equation}\label{KrevoA}
  \frac{d}{dt}g^{\lambda\mu} (t) = \hat{\sf k}^{\lambda\mu}(t) =\sum_{\alpha\beta}\left(
  \sum_{\gamma\delta}{\sf k}^{\gamma\delta}(t) \Pi^\lambda_{\gamma\alpha}\,\overline{\Pi^\mu_{\delta\beta}}
  \right)\,g^{\alpha\beta}(t) =: \sum_{\alpha\beta} A_{\alpha\beta}^{\lambda\mu}(t)\,g^{\alpha\beta}(t)
  \;.
\end{equation}
This is the result we were looking for, which is a differential equation for the GKS matrix $g(t)$ of ${\mathcal V}(t)$
in terms of the GKS matrix ${\sf k}(t)$ of ${\mathcal K}(t)$ and some constant array of complex numbers that are, in principle, known.
Note that the differential equation (\ref{KrevoA}) comprises the total time evolution for any initial value $\rho_0$.

\subsection{Time-dependent Lindblad equation}\label{sec:TDLE}
In this subsection we will clarify the connection between the considerations of the preceding subsection \ref{sec:TDGKS}
and the usual approach using the time-dependent Lindblad equation, see \cite{GKS76} and \cite{B12}.
First we will recapitulate the derivation of the Lindblad equation.

One assumes that the object system with Hilbert space ${\mathcal H}$ is coupled to another system (``environment")
with Hilbert space ${\mathcal H}_E$ such that the time evolution of the total system is given by a unitary
operator $U(t,0)$ satisfying $U(0,0)={\mathbbm 1}$. Initially, at time $t=0$, the state of the total system is given by the mixed product state
$\rho_0\otimes \rho_E\in {\sf B}_1^{+}\left({\mathcal H}\otimes {\mathcal H}_E\right)$. If we reduce to the system's
state at time $t$ by performing the partial trace $\mbox{Tr}_E$ we obtain
\begin{equation}\label{rhotA}
  \rho(t)={\mathcal V}(t)\,\rho_0 := \mbox{Tr}_E \left( U(t,0)\,\rho_0\otimes \rho_E U(t,0)^\ast\right)
  \;.
\end{equation}

It can be shown that the superoperator ${\mathcal V}(t)$ defined by (\ref{rhotA}) is a linear, trace-preserving
and completely positive operator ${\mathcal V}(t)\in L({\mathcal A})$. Hence (\ref{evolution}) holds and,
moreover, the derivative $\frac{d}{dt}\rho(t)$ is given by
\begin{equation}\label{KrepA}
 \frac{d}{dt} \rho(t) = {\mathcal K}(t)  \rho(t) = \sum_{0\le\alpha,\beta<N^2}{\sf k}^{\alpha\beta}(t)\, F_\alpha\,\rho(t)\,F_\beta^\ast
 \;,
\end{equation}
where $\left(F_\alpha\right)_{0\le \alpha<N^2}$ is a fixed orthonormal basis in ${\mathcal A}$,
and the ${\sf k}^{\alpha\beta}(t)$ are the entries of an Hermitean matrix ${\sf k}(t) \in{\sf B}\left({\mathbb C}^{N\times N}\right)$,
see (\ref{Krep}).

In the next step one specializes the basis $\left(F_\alpha\right)_{0\le \alpha<N^2}$ by setting
$F_0={\mathbbm 1}/\sqrt{N}$. According to
\begin{equation}\label{Falphatrace}
 \delta_{0 \alpha}=\langle F_0|F_\alpha \rangle=\mbox{Tr }\left( {\mathbbm 1}/\sqrt{N} F_\alpha \right)
\end{equation}
this implies that the remaining basis operators $\left(F_\alpha\right)_{1\le \alpha<N^2}$
have vanishing trace. Splitting the double sum in (\ref{KrepA}) according to whether $\alpha,\beta=0$  or $>0$ we obtain
\begin{equation}\label{KrepB}
{\mathcal K}(t)  \rho(t) = {\sf k}^{00}(t)/N\,\rho(t) +\sum_{1\le \alpha<N^2} {\sf k}^{\alpha 0}(t)/\sqrt{N}\,F_\alpha\,\rho(t)
 +\sum_{1\le \beta<N^2} {\sf k}^{0 \beta }(t)/\sqrt{N}\,\rho(t)\,F_\beta^\ast
+ \sum_{1\le\alpha,\beta<N^2}{\sf k}^{\alpha\beta}(t)\, F_\alpha\,\rho(t)\,F_\beta^\ast
 \;.
\end{equation}
Upon defining
\begin{eqnarray}
\label{defFGHa}
  F(t) &:=& \sum_{1\le \alpha<N^2} {\sf k}^{\alpha 0}(t)F_\alpha=: \widetilde{G(t)}-{\sf i}H(t),\; \widetilde{G(t)},H(t)\in {\sf B}({\mathcal H}),\\
  \label{defFGHb}
  G(t) &:=& \widetilde{G(t)}+ {\sf k}^{00}(t)/(2N)\, {\mathbbm 1}
  \;,
\end{eqnarray}
one obtains, after a short calculation,
\begin{equation}\label{KrepC}
 {\mathcal K}(t)  \rho(t) = -{\sf i}[H(t),\rho(t)]+\{ G(t),\rho(t)\}+\sum_{1\le\alpha,\beta<N^2}{\sf k}^{\alpha\beta}(t)\, F_\alpha\,\rho(t)\,F_\beta^\ast
 \;.
\end{equation}
The condition that ${\mathcal K}(t)$ is trace-annihilating implies
\begin{equation}\label{annihi}
  0=\mbox{Tr } \left( {\mathcal K}\,\rho\right) = \mbox{Tr } \left( \underbrace{\left(
  2G+\sum_{1\le \alpha,\beta<N^2}{\sf k}^{\alpha\beta}\,F_\beta^\ast F_\alpha
   \right)}_0 \,\rho\right)\quad\mbox{for all }\rho\in{\sf B}({\mathcal H})
\end{equation}
and hence
\begin{equation}\label{annihiG}
 G(t)=-\frac{1}{2}\sum_{1\le \alpha,\beta<N^2}{\sf k}^{\alpha\beta}(t)\,F_\beta^\ast F_\alpha
 \;.
\end{equation}
Then (\ref{KrepC}) can be written in the form
\begin{equation}\label{KrepD}
 {\mathcal K}(t)  \rho(t) = -{\sf i}[H(t),\rho(t)]
 +\sum_{1\le\alpha,\beta<N^2}{\sf k}^{\alpha\beta}(t)\left( F_\alpha\,\rho(t)\,F_\beta^\ast
 -\frac{1}{2}\left\{ F_\beta^\ast F_\alpha,\rho(t)\right\}\right)
 \;,
\end{equation}
using the anti-commutator $\{A,B\}:=AB+BA$ of Hermitean operators. 
If one makes a Markov approximation for the time evolution of the open system, 
then one can regard ${\mathcal K}(0)$ as the generator of a dynamic semigroup, 
cf.~equation (2.3) in \cite{GKS76} or (7) in \cite{Betal01}.
In this context, $k^{\alpha \beta}(0)$ is referred to as the ``GKS matrix," 
see \cite{Betal01}, which is related to our notation but must nevertheless be distinguished.

Next one diagonalizes the Hermitean matrix ${\sf k}\in L({\mathbb C}^{N^2-1})$ and writes
\begin{equation}\label{diagk}
  {\sf k}(t) = W(t)\,\Delta(t)\,W(t)^\ast
  \;,
\end{equation}
where the diagonal matrix $\Delta(t)$ contains the real eigenvalues ${\sf k}^\alpha(t)$ of ${\sf k}(t)$
and the columns of the unitary matrix $W(t)$ are formed of the corresponding eigenvectors.
Introducing the time-dependent Lindblad operators
\begin{equation}\label{defLind}
 A_\alpha(t):= \sum_{1\le \beta <N^2} W_{\beta \alpha}(t) F_\beta\;,
\end{equation}
which have, by construction, vanishing trace, we finally obtain the usual form of the Lindblad equation
\begin{equation}\label{KrepE}
 {\mathcal K}(t)  \rho(t) = -{\sf i}[H(t),\rho(t)]
 +\sum_{1\le\alpha<N^2}{\sf k}^{\alpha}(t)\left( A_\alpha(t)\,\rho(t)\,A_\alpha^\ast(t)
 -\frac{1}{2}\left\{ A_\alpha^\ast(t) A_\alpha(t),\rho(t)\right\}\right)
 \;.
\end{equation}

We will show that (\ref{KrepE}) is equivalent to the evolution equation (\ref{KrepA})
which is identical to (\ref{Krep}) considered in the
subsection \ref{sec:TDGKS}. Having just recapitulated the well-known derivation  (\ref{KrepA}) $\Rightarrow$  (\ref{KrepE}),
we now consider the reverse case (\ref{KrepE}) $\Rightarrow$  (\ref{KrepA}).

First note that (\ref{KrepE}) presupposes the orthonormal basis  $\left( A_1(t),\ldots, A_{N^2-1}(t)\right)$
in ${\mathcal A}_0$, the subspace of $L({\mathcal H})$  of operators with vanishing trace.
Let  $\left(F_\alpha\right)_{1\le \alpha<N^2}$ be another fixed orthonormal basis in ${\mathcal A}_0$, then
there exists a unitary matrix $W(t)\in L({\mathbbm C}^{N^2-1})$ such that
\begin{equation}\label{unitaryFA}
 A_\alpha(t)=\sum_{1\le \beta <N^2}W_{\beta\alpha}(t)\,F_\beta,\quad \mbox{for all } 1\le \alpha<N^2
 \;.
\end{equation}
This allows the reconstruction of the Hermitean matrix
\begin{equation}\label{reconk}
 {\sf k}(t):=W(t)\Delta(t)\,W(t) \in {\sf B}({\mathbbm C}^{N^2-1})
 \;,
\end{equation}
since the diagonal matrix $\Delta(t)$ contains the given eigenvalues $\left({\sf k}^\alpha(t)\right)_{1\le\alpha<N^2}$.
The remaining entries of the full matrix ${\sf k}(t) \in {\sf B}({\mathbbm C}^{N^2})$ are reconstructed
as follows. ${\sf k}^{00}(t)$ is  uniquely determined by the condition $\mbox{Tr }{\sf k}(t)=0$, see (\ref{traceanniac}).
From $H(t)$ and $G(t)$ we obtain $F(t)$ via (\ref{defFGHa}) and (\ref{defFGHb}).
The expansion of $F(t)$ into the basis $\left(F_\alpha\right)_{1\le \alpha<N^2}$ gives
\begin{equation}\label{expandF}
 F=\sum_{1\le\alpha<N^2}{\sf k}^{\alpha 0}(t)/\sqrt{N}\,F_\alpha
 \;,
\end{equation}
and hence the entries ${\sf k}^{\alpha 0}(t)$ and ${\sf k}^{0\alpha} (t)=\overline{{\sf k}^{\alpha 0}(t)}$.
Thus we obtain   (\ref{KrepA}) for the orthonormal basis $\left(F_0:={\mathbbm 1}/\sqrt{N},F_1,\ldots, F_{N^2-1}\right)$.
The transformation to another arbitrary orthonormal basis in ${\mathcal A}$ can be performed and gives  (\ref{KrepA})
with a correspondingly transformed GKS-matrix analogously to (\ref{transGKSmatrixa}) - (\ref{transGKSmatrixd}).

\subsection{Expansion up to $O(t^2)$ terms}\label{sec:EXP}

 Examples in which the time evolution of the total system can be calculated exactly are rare.
 Instead, we want to illustrate the above considerations with a series expansion w.~r.~t.~time $t$,
 which applies to general open quantum systems.
It is possible to calculate up to the second order in $t$ with reasonable effort.
We will calculate the GKS-matrix $g(t)$ of the time evolution map ${\mathcal V}(t)$ and confirm that its eigenvalues are non-negative,
as it must be since the time evolution is completely positive.

We make the general assumptions of Section \ref{sec:TDLE} and hence consider a time evolution of
the open quantum system given by eq.~(\ref{rhotA}). The orthonormal basis $\left(F_\alpha \right)_{0\le \alpha<N^2}$
of ${\mathcal A}$ is assumed to satisfy $F_0={\mathbbm 1}/\sqrt{N}$ and, to further simplify the calculations, to consist of Hermitean operators,
i.~e., $F_\alpha\in {\sf B}_0({\mathcal H})$ for $1\le \alpha<N^2$.
Concerning the environment, we additionally assume that its Hilbert space
is finite-dimensional, say, $\mbox{dim }({\mathcal H}_E)=M$. Moreover, we will restrict ourselves to the case
where $\rho_E$ is a pure state. After introducing a suitable orthonormal base $\left( |i\rangle\right)_{0\le i <M}$ in ${\mathcal H}_E$
we may write
\begin{equation}\label{defrhoE}
 \rho_E=|0\rangle\langle 0|
 \;.
\end{equation}
It will be convenient to also choose the associated orthonormal basis $\left(E_\gamma \right)_{0\le \gamma\le M^2}$ in $L({\mathcal H}_E)$
defined analogously as in subsection \ref{sec:GKSCI}. We will re-order this basis writing it in the way
 $\left(E_\gamma \right)_{-\lfloor M^2/2\rfloor \le \gamma \le \lfloor M^2/2\rfloor }$ for odd $M$
 (for even $M$ one $\le $ sign has to be replaced by $<$ in the following equations).
Moreover, we choose
\begin{equation}\label{deforderE}
 E_0= \rho_E=|0\rangle\langle 0|,\, E_\gamma= |\gamma\rangle \langle 0|\quad\mbox{for } \gamma=1,\ldots, M-1
 ,\quad\mbox{and } E_{-\gamma} = E_\gamma^\ast=|0\rangle \langle\gamma| \mbox{ for all } -\lfloor M^2/2\rfloor \le  \gamma \le \lfloor M^2/2\rfloor
 \;.
\end{equation}

This implies
\begin{equation}\label{TrGrho}
 \mbox{Tr}_E \left(E_\gamma\,\rho_E \right)= \left\{
 \begin{array}{l@{\quad:\quad}r}
   \mbox{Tr}_E |\gamma\rangle \langle 0|0\rangle \langle 0|=\delta_{\gamma 0} & 0\le \gamma < M \\
   0 & \mbox{else}
 \end{array}
 \right.
 \;,
\end{equation}
and
\begin{eqnarray}\label{TrGrhoGp}
 \mbox{Tr}_E\left(E_\gamma\,\rho_E\,E_\delta^\ast \right)&=&
 \left\{
 \begin{array}{l@{\quad:\quad}r}
  \mbox{Tr}_E\,|\gamma\rangle \langle 0|0\rangle  \langle 0|0\rangle\langle \delta|=\delta_{\gamma\delta} & 0\le \gamma,\delta < M \\
   0 & \mbox{else}
 \end{array}
 \right.
 \;,\\
 \label{TrGrhoGn}
 \mbox{Tr}_E\left(E_\delta^\ast\,\rho_E\,E_\gamma \right)&=&
 \left\{
 \begin{array}{l@{\quad:\quad}r}
  \mbox{Tr}_E\,|\delta\rangle \langle 0|0\rangle  \langle 0|0\rangle\langle \gamma|=\delta_{\gamma\delta} & -M < \gamma,\delta \le 0 \\
   0 & \mbox{else}
 \end{array}
 \right.
 \;.
\end{eqnarray}

The total time evolution $U(t,0)$ will be generated by a, generally time-dependent,
Hamiltonian $K(t)\in {\sf B}_0({\mathcal H}\otimes{\mathcal H}_E)$ such that
\begin{equation}\label{UKt}
 \frac{d}{dt}U(t,0) = -{\sf i}\, K(t)\,U(t,0)
 \;.
\end{equation}
$K(t)$ can be expanded into the product basis $\left( F_\alpha\otimes E_\gamma\right)_{0\le \alpha<N^2,\,{-\lfloor M^2/2\rfloor \le  \gamma \le \lfloor M^2/2\rfloor}}$
with the result
\begin{eqnarray}\label{Kexpand1}
 K(t)&=&H_S(t)\otimes {\mathbbm 1}_E +{\mathbbm 1}\otimes H_E(t) +H_{SE}(t),\\
 \label{Kexpand2}
 H_S(t) &=& \sum_{1\le \alpha<N^2} h^{\alpha}(t)\,F_\alpha,\\
 \label{Kexpand3}
 H_E(t) &=& \sum_{-\lfloor M^2/2\rfloor \le  \gamma \le \lfloor M^2/2\rfloor } h_E^{\gamma}(t)\,E_\gamma,\\
 \label{Kexpand4}
 H_{SE}(t)&=&\sum_{1\le \alpha<N^2\atop -\lfloor M^2/2\rfloor \le  \gamma \le  \lfloor M^2/2\rfloor } v^{\alpha \gamma}(t)\,F_\alpha\otimes E_\gamma
 \;,
\end{eqnarray}
such that the functions $h^\alpha(t)$  are always real-valued, but  $v^{\alpha\gamma}(t)$ and $h_E^\gamma(t)$ may be complex.
Especially,  $v^{\alpha,\gamma}(t)$ satisfies
\begin{equation}\label{condv}
 v^{\alpha,\gamma}(t) =\overline{v^{\alpha,-\gamma}(t)},\quad \mbox{for all } 1\le \alpha<N^2 \mbox{ and } -\lfloor M^2/2\rfloor \le  \gamma\le \lfloor M^2/2\rfloor
 \;,
\end{equation}
which implies
\begin{equation}\label{condv0}
 v^{\alpha,0}(t)\in {\mathbb R},\quad \mbox{for all } 1\le \alpha<N^2
 \;.
\end{equation}

In what follows summations over $\alpha,\beta,\ldots$ will always run from $1$ to $N^2-1$
and over $\gamma,\delta,\ldots$  from $-\lfloor M^2/2\rfloor$ to $\lfloor M^2/2\rfloor$
if not indicated otherwise.
Up to terms of order $O(t^2)$ we have
\begin{equation}\label{drhoexp}
 \rho(t)= \rho_0 + \left.\frac{d}{dt}\rho(t)\right|_{t=0}\,t + \left.\frac{d^2}{dt^2}\rho(t)\right|_{t=0}\,\frac{t^2}{2} +O(t^3)
 \;.
\end{equation}
The time derivative of (\ref{rhotA}) is given by
\begin{equation}\label{drhotA}
  \frac{d}{dt}\rho(t)=-{\sf i}\,\mbox{Tr}_E \left[ K(t),U(t,0)\rho_0\otimes \rho_E U(t,0)^\ast
  \right]
  \;,
\end{equation}
and hence
\begin{equation}\label{drhotA0}
  \left.\frac{d}{dt}\rho(t)\right|_{t=0}=-{\sf i}\,\mbox{Tr}_E \left[ K(0),\rho_0\otimes \rho_E  \right]
  \;,
\end{equation}
and
\begin{equation}\label{ddrhotA0}
  \left.\frac{d^2}{dt^2}\rho(t)\right|_{t=0}=-{\sf i}\,\left.\mbox{Tr}_E \left[ \frac{d}{dt}K(t)\right|_{t=0},\rho_0\otimes \rho_E  \right]
  -\,\mbox{Tr}_E \left[ K(0),\left[K(0),\rho_0\otimes \rho_E  \right]\right]
  \;.
\end{equation}

The series expansion of the corresponding GKS-matrix $g(t)$ will be denoted as
\begin{equation}\label{serg}
g(t)=g_0 + g_1\, t + g_2\, t^2 +O(t^3)
\;.
\end{equation}

\subsubsection{$0$th order in $t$}

First consider the $0$th order. Here we have
\begin{equation}\label{zero}
  {\mathcal V}(0)\,\rho_0=\rho_0 = N\,F_0\,\rho_0\,F_0
  \;,
\end{equation}
and hence the  initial GKS-matrix $g_0^{\alpha,\beta}$ has the only non-vanishing entry $g_0^{00}=N$.
Its eigenvalues are hence $N,0$, where the eigenvalue $0$ is $(N^2-1)$-fold degenerate.

\subsubsection{$1$st order in $t$}

Then we consider the first order  w.~r.~t.~$t$ and calculate the expressions that result
when the three terms of the Hamiltonian (\ref{Kexpand1}) are inserted into (\ref{drhotA0}),
whereby the possible time dependence of the occurring terms is suppressed.
\begin{equation}\label{term1}
  -{\sf i}\,\mbox{Tr}_E \left[ H_S \otimes {\mathbbm 1}_E, \rho_0\otimes \rho_E\right]=   -{\sf i}\, \left[H_S,\rho_0 \right]
  =  -{\sf i}\, \sum_\alpha h^\alpha \left( F_\alpha\,\rho_0 - \rho_0\,F_\alpha\right)
  \;.
\end{equation}

\begin{equation}\label{term2}
  -{\sf i}\,\mbox{Tr}_E \left[ {\mathbbm 1} \otimes H_E, \rho_0\otimes \rho_E\right]=   -{\sf i} \,\mbox{Tr}_E
  \rho_0 \otimes  \left[ H_E,\rho_E\right] =   -{\sf i}\,\rho_0\, \mbox{Tr}_E  \left[ H_E,\rho_E\right] =0
  \;.
\end{equation}

\begin{eqnarray}\label{term31a}
  -{\sf i}\,\mbox{Tr}_E \left[H_{SE}, \rho_0\otimes \rho_E\right]
  &=&
   -{\sf i} \,\mbox{Tr}_E
  \sum_{\alpha \gamma}v^{\alpha \gamma}\,\left[ F_\alpha \otimes E_\gamma, \rho_0\otimes \rho_E\right]\\
  \label{term31b}
 &=&   -{\sf i} \,\mbox{Tr}_E
  \sum_{\alpha \gamma}v^{\alpha \gamma}\,\left(F_\alpha \rho_0\otimes E_\gamma \rho_E-\rho_0 F_\alpha\otimes \rho_E E_\gamma \right)\\
  \label{term31c}
 &=&  -{\sf i} \,\sum_{\alpha\gamma} v^{\alpha \gamma}\,\left[ F_\alpha, \rho_0\right]\,\mbox{Tr}_E\left(E_\gamma \rho_E \right)\\
 \label{term31d}
 &\stackrel{(\ref{TrGrho})}{=}& -{\sf i}\, \sum_{\alpha}v^{\alpha 0}\,\left( F_\alpha\, \rho_0-\rho_0\,F_\alpha\right)
  \;.
\end{eqnarray}

It follows that the $O(t)$ corrections  $g_1^{\alpha,\beta}$ to the GKS-matrix have only non-zero contributions
in the $0$th row or $0$th column, i.~e., for $g_1^{0\alpha}$ and $g_1^{\alpha 0}$:
\begin{equation}\label{g1}
 g_1=\sqrt{N} \left(
 \begin{array}{cccc}
   0 & {\sf i}(h^1+v^{1,0}) & \ldots & {\sf i}(h^{N^2-1}+v^{N^2-1,0}) \\
   -{\sf i}(h^1+v^{1,0}) & 0 & \ldots & 0 \\
   \vdots & 0 & \ldots & 0 \\
  -{\sf i}(h^{N^2-1}+v^{N^2-1,0}) &0 & \ldots & 0
 \end{array}
 \right)
 \;.
\end{equation}
Hence the corrections to the eigenvalues $N,0$ of $g(0)$ are of order $O(t^2)$ and have to be
disregarded in linear order w.r.t.~$t$, see below for the details of the perturbational calculation.

\subsubsection{$2$nd order in $t$}

For the second order in $t$ we will focus on the contributions to the GKS matrix for $\alpha,\beta \ge 1$.
The contributions for $\alpha=0$ or $\beta=0$ will give third order corrections to the degenerate eigenvalue $0$
and second order corrections to the dominant eigenvalue $N$. The latter does not jeopardize the positivity of the
dominant eigenvalue as long as $t$ remains small.
For details, we refer the reader to the following perturbation-theoretical calculations.

Proceeding analogously as in the case of linear order we would have to calculate nine terms which
result from inserting the three terms of the Hamiltonian (\ref{Kexpand1}) into (\ref{ddrhotA0}).
Here we will confine ourselves to present only the non-vanishing results. The omission of terms
that give only contributions to the $0$th row or $0$th column of $g_2$
will be indicated by the symbol $\stackrel{\alpha,\beta\ge 1}{\longrightarrow}$. These terms,
except the contribution to $g_2^{00}$,
give corrections to the eigenvalues of $g(t)$ of maximal order $O(t^3)$ and hence can be dismissed.
The $g_2^{00}$-term can be disregarded since it will only yield a $O(t^2)$-correction to the dominant eigenvalue
$N$  of $g(t)$ which hence remains positive.

\begin{eqnarray}
\label{term11a}
 -{\mbox Tr}_E\left[H_S\otimes {\mathbbm 1}_E,\left[H_S\otimes {\mathbbm 1}_E,\rho_0\otimes \rho_E \right] \right]&=&
 -{\mbox Tr}_E \sum_{\alpha\beta}h^\alpha\,h^\beta\,\left[ F_\beta,\left[F_\alpha,\rho_0\right]\right]\otimes\rho_E=
 -\sum_{\alpha\beta}h^\alpha\,h^\beta\,\left[ F_\beta,\left[F_\alpha,\rho_0\right]\right]\\
 &=& -\sum_{\alpha\beta}h^\alpha\,h^\beta\,
 \left(F_\beta\left(F_\alpha\rho_0-\rho_0 F_\alpha \right)- \left(F_\alpha\rho_0-\rho_0 F_\alpha \right)F_\beta\right)\\
 \label{term11b}
 &\stackrel{\alpha,\beta\ge 1}{\longrightarrow}& \sum_{\alpha\beta}h^\alpha\,h^\beta\,
 \left( F_\beta\rho_0 F_\alpha + F_\alpha\rho_0 F_\beta \right)=2\,\sum_{\alpha\beta}h^\alpha\,h^\beta\, F_\alpha\rho_0 F_\beta
 \;.
\end{eqnarray}

\begin{eqnarray}
\label{term31a}
 -\mbox{Tr}_E\left[H_{SE},\left[H_S\otimes {\mathbbm 1}_E,\rho_0\otimes \rho_E \right] \right]&=&
  -\mbox{Tr}_E \sum_{\alpha\beta\gamma}v^{\alpha\gamma}\,h^\beta\,\left[F_\alpha\otimes E_\gamma,
  \left[ F_\beta,\rho_0   \right]\otimes \rho_E\right]\\
  \label{term31b}
  &=& -\mbox{Tr}_E \sum_{\alpha\beta\gamma}v^{\alpha\gamma}\,h^\beta\,
  \left(
  F_\alpha\left[ F_\beta,\rho_0\right]\otimes E_\gamma\,\rho_E-
  \left[F_\beta,\rho_0\right]F_\alpha\otimes \rho_E\, E_\gamma
  \right)\\
  \label{term31c}
  &=&-\sum_{\alpha\beta\gamma}v^{\alpha\gamma}\,h^\beta\,
  \left(
  F_\alpha\left[ F_\beta,\rho_0\right]-
  \left[ F_\beta,\rho_0\right]F_\alpha\right)
  \,
  \mbox{Tr}_E \left( E_\gamma\,\rho_E\right)\\
  \label{term31d}
  &\stackrel{(\ref{TrGrho})}{=}&
  -\sum_{\alpha\beta}v^{\alpha 0}\,h^\beta\,
 \left[
  F_\alpha,\left[ F_\beta,\rho_0\right]\right)\\
 \label{term31e}
 &=&
  -\sum_{\alpha\beta}v^{\alpha 0}\,h^\beta\,
  \left(
  F_\alpha F_\beta \rho_0-F_\alpha\rho_0 F_\beta-F_\beta\rho_0 F_\alpha +\rho_0 F_\beta F_\alpha
  \right)\\
   \label{term31f}
  &\stackrel{\alpha,\beta\ge 1}{\longrightarrow}&
  \sum_{\alpha\beta}\left( v^{\alpha 0}\,h^\beta+v^{\beta 0}\,h^\alpha\right) F_\alpha \rho_0 F_\beta
  \;.
\end{eqnarray}
An analogous calculation yields
\begin{equation}\label{term13}
  -\mbox{Tr}_E\left[H_S\otimes {\mathbbm 1}_E,\left[H_{SE},\rho_0\otimes \rho_E \right] \right]
  \stackrel{\alpha,\beta\ge 1}{\longrightarrow}
  \sum_{\alpha\beta} \left( v^{\alpha 0}\,h^\beta+v^{\beta 0}\,h^\alpha\right) F_\alpha \rho_0 F_\beta
  \;,
\end{equation}
the same result as (\ref{term31f}).
The last non-vanishing contribution to $g(t)$ is given by
\begin{eqnarray}
\label{term33a}
  -\mbox{Tr}_E\left[H_{SE},\left[H_{SE},\rho_0\otimes \rho_E \right] \right]&=&
   -\mbox{Tr}_E\left[H_{SE},\left[H_{SE}^\ast,\rho_0\otimes \rho_E \right] \right]
  \\
  \label{term33ab}
  &=&
   -\mbox{Tr}_E \sum_{\alpha\beta\gamma\delta} v^{\alpha\gamma}\,\overline{v^{\beta\delta}}\,
   \left[F_\alpha\otimes E_\gamma,\left[F_\beta\otimes E_\delta^\ast,\rho_0\otimes \rho_E\right] \right]\\
   \label{term33b}
   &=& -\mbox{Tr}_E \sum_{\alpha\beta\gamma\delta} v^{\alpha\gamma}\,\overline{v^{\beta\delta}}\,
   \left(F_\alpha F_\beta\rho_0 \otimes E_\gamma E_\delta^\ast \rho_E-
   F_\beta \rho_0 F_\alpha \otimes E_\delta^\ast  \rho_E E_\gamma
   \right.\\
   \nonumber
   &&\quad\quad \quad \quad \left.
   -F_\alpha \rho_0 F_\beta \otimes E_\gamma\rho_E E_\delta^\ast +\rho_0 F_\beta F_\alpha\otimes \rho_E E_\delta^\ast E_\gamma
   \right)\\
   \label{term33c}
    &\stackrel{\alpha,\beta\ge 1}{\longrightarrow}&
    \sum_{\alpha\beta\gamma\delta} v^{\alpha\gamma}\,\overline{v^{\beta\delta}}\,
    \left(
    F_\beta \rho_0 F_\alpha \mbox{Tr}_E\left( E_\delta^\ast \rho_E E_\gamma\right)
    +
    F_\alpha \rho_0 F_\beta  \mbox{Tr}_E\left( E_\gamma \rho_E E_\delta^\ast\right)
    \right)\\
    \label{term33d}
     &\stackrel{(\ref{TrGrhoGp},\ref{TrGrhoGn})}{=}&
      \sum_{\alpha\beta}\sum_{-M<\gamma\le 0} v^{\alpha\gamma}\,\overline{v^{\beta\gamma}}\,F_\beta \rho_0 F_\alpha
      +
       \sum_{\alpha\beta}\sum_{0\le\gamma<M} v^{\alpha\gamma}\,\overline{v^{\beta\gamma}}\,F_\alpha \rho_0 F_\beta
    \\
    \label{term33e}
    &=&
     \sum_{\alpha\beta}\sum_{0\le\gamma <M} v^{\beta,-\gamma}\,\overline{v^{\alpha,-\gamma}}\,F_\alpha \rho_0 F_\beta
      +
       \sum_{\alpha\beta}\sum_{0\le\gamma<M} v^{\alpha\gamma}\,\overline{v^{\beta\gamma}}\,F_\alpha \rho_0 F_\beta\\
    &\stackrel{(\ref{condv})}{=}& 2 \sum_{\alpha\beta}\sum_{0\le\gamma<M} v^{\alpha\gamma}\,\overline{v^{\beta\gamma}}\,F_\alpha \rho_0 F_\beta\\
     &\stackrel{(\ref{condv0})}{=}& 2 \sum_{\alpha\beta} v^{\alpha 0}\,{v^{\beta 0}}\,F_\alpha \rho_0 F_\beta
         +
         2 \sum_{\alpha\beta}\sum_{1\le\gamma<M} v^{\alpha\gamma}\,\overline{v^{\beta\gamma}}\,F_\alpha \rho_0 F_\beta
         \;.
   \end{eqnarray}

Summarizing, we obtain the second order expansion of the submatrix $\left(g^{\alpha\beta}(t)\right)_{1\le \alpha,\beta<N^2}$:
\begin{equation}\label{subg2}
g^{\alpha\beta}(t) =g_2^{\alpha\beta}\,t^2 +O(t^3)=\left( \left( h^\alpha +v^{\alpha 0}\right)\ \left( h^\beta +v^{\beta 0}\right)
+\sum_{1\le\gamma<M} v^{\alpha\gamma}\,\overline{v^{\beta\gamma}}
\right)\,t^2+O(t^3)
\;,
\end{equation}
which is obviously positively semi-definite.
The latter is a necessary condition for the full matrix $\left(g^{\alpha\beta}(t)\right)_{0\le \alpha,\beta<N^2}$
being  positively semi-definite, but, in general, not a sufficient one. However, in the present case
the validity of
\begin{equation}\label{gtpos}
 g_0+ g_1 \,t + g_2\,t^2 \ge 0
\end{equation}
can be shown by means of perturbational calculations.

To this end we write the general eigenvalue equation $g(t)\varphi(t)=\epsilon(t)\varphi(t)$ in the form
\begin{equation}\label{expandeigenvalue}
  \left( g_0+g_1\,t+g_2\,t^2 +\ldots\right)\left( \varphi
  _0+\varphi_1\,t+\varphi_2\,t^2+\ldots\right)
  =
  \left(\epsilon_0+\epsilon_1\,t+\epsilon_2\,t^2+\ldots\right)\left( \varphi_0+\varphi_1\,t+\varphi_2\,t^2+\ldots\right)
\end{equation}
and expand it into powers of $t$. A usual, we employ a normalization condition of the form
\begin{equation}\label{norm}
1= \left\langle \varphi_0\left| \right. \varphi\right\rangle =  \underbrace{\left\langle \varphi_0\left| \right. \varphi_0\right\rangle}_1
+t\,\underbrace{\left\langle \varphi_0\left| \right. \varphi_1\right\rangle}_0 +
 t^2\,\underbrace{\left\langle \varphi_0\left| \right. \varphi_2\right\rangle}_0+\ldots
 \;,
\end{equation}
which implies $ \varphi_0\perp\varphi_1,\varphi_2,\ldots$.

\subsubsection{$t^0$-terms}
We conclude
\begin{equation}\label{eigent0}
  g_0\,\varphi_0=\epsilon_0\,\varphi_0
  \;,
\end{equation}
and that, according to what has been said before, either $\epsilon_0=N$ and $\varphi_0=(1,0,\ldots,0)^\top$
or $\epsilon_0=0$ and $\varphi_0 \perp (1,0,\ldots,0)^\top$.
\subsubsection{$t^1$-terms}
We obtain
\begin{equation}\label{eigent1}
  g_0\,\varphi_1+ g_1\,\varphi_0=\epsilon_0\,\varphi_1+\epsilon_1\,\varphi_0
  \;,
\end{equation}
and, upon forming the inner product with $\varphi_0$ and using (\ref{norm}), the familiar result
\begin{equation}\label{eigent1a}
  \epsilon_1 = \left\langle \varphi_0\left| g_1\right| \varphi_0\right\rangle
    \;.
\end{equation}
For $\epsilon_0=N$ and $\varphi_0=(1,0,\ldots,0)^\top$ we conclude $\epsilon_1=g_1^{00}=0$.
Now consider the case $\epsilon_0=0$ and $\varphi_0\perp(1,0,\ldots,0)^\top$ and let $\left(\varphi_0^{(i)}\right)_{0\le i <N^2}$
be an orthonormal eigenbasis of $g_0$ such that $\varphi_0^{(0)}=\varphi_0$.
Then the inner product of (\ref{eigent1}) with $\varphi_0^{(i)}$
gives
\begin{equation}\label{eigent1b}
  \left\langle \varphi_0^{(i)} \left| g_1\right| \varphi_0\right\rangle =0,\quad \mbox{ for }0<i<N^2
  \;.
\end{equation}
According to (\ref{g1}) $g_1\,\varphi_0$ is of the form $g_1\,\varphi_0=(x,0,\ldots,0)^\top\perp \varphi_0$
and thus, also in this case,  $\epsilon_1=0$. All corrections to the eigenvalues of $g_0$ are at most of order $O(t^2)$.

\subsubsection{$t^2$-terms}

We obtain
\begin{equation}\label{eigent2}
  g_0\,\varphi_2+ g_1\,\varphi_1+ g_2\,\varphi_0=\epsilon_0\,\varphi_2+\epsilon_1\,\varphi_1+\epsilon_2\,\varphi_0
  \;,
\end{equation}
and, upon forming the inner product with $\varphi_0$ and using (\ref{norm}),
\begin{equation}\label{eigent2a}
 \epsilon_2=\left\langle \varphi_0 \left| g_1 \right| \varphi_1\right\rangle+\left\langle \varphi_0 \left| g_2 \right| \varphi_0\right\rangle
 \;.
\end{equation}
For the case of $\epsilon_0=0$ we may expand $\varphi_1\perp \varphi_0$ into the eigenbasis $\left(\varphi_0^{(i)}\right)_{0\le i <N^2}$
and, using (\ref{eigent1b}), conclude
\begin{equation}\label{eigent2b}
 \epsilon_2=\left\langle \varphi_0 \left| g_2 \right| \varphi_0\right\rangle
 \;.
\end{equation}
It follows that  $\epsilon_2\ge 0$ since $\varphi_0\perp (1,0,\ldots,0)^\top$ and the submatrix
$\left(g_2^{\alpha\beta}\right)_{1\le \alpha,\beta<N^2}$  is positively semi-definite.

Although for $\epsilon_0=N$  the  correction $\epsilon_2\,t^2$ will be, in general, negative, this is only a correction
of order $O(t^2)$ and hence the dominant eigenvalue  will remain positive.

This completes our arguments showing that the GKS-matrix $g(t)$ corresponding to the time evolution of an open
quantum system will be positively semi-definite if we consider the expansion $g(t)= g_0+ g_1 \,t + g_2\,t^2 +O(t^3)$
up to second order in $t$. Recall that a non-vanishing $2$nd order contribution $g_2\,t^2$  to the GKS-matrix
corresponds to the case where the semi-group  property of the time evolution is violated.
Moreover, the restriction to a pure initial state $\rho_E(0)$ of the environment is harmless in this respect, since the
GKS-matrix $g(t)$ depends linearly on $\rho_E(0)$. If $g_0+ g_1 \,t + g_2\,t^2 \ge 0$ for pure $\rho_E(0)$ then
this also holds for all general mixed statistical operators $\rho_E(0)$.

\section{Summary}\label{sec:SUM}

Even 70 years after its definition, it seems that the concept of a ``completely positive transformation"
is still not fully understood mathematically and physically.
The same applies to the criterion for complete positivity, which is provided by the Choi isomorphism.
This isomorphism depends on the choice of an orthonormal basis $\left(| i\rangle\right)_{0\le i<N}$ in Hilbert space.
An obvious motive in the search for generalizations is therefore the extension of bases
of the form $\left(| i\rangle\langle j|\right)_{0\le i,j<N}$ to arbitrary
orthonormal bases $\left( F_\alpha\right)_{0\le \alpha <N^2}$ in the space of complex $N\times N$-matrices.
Such an extension had already existed for some time in the form of the dePillis-Jamio{\l}ski
isomorphism, but the criterion for complete positivity had to be sacrificed for it.
In the articles \cite{PS13} and \cite{FC24}, however, this problem has been solved.

In the present work we have followed a different approach which can be traced back
to the seminal paper \cite{GKS76} by Gorini, Kossakowski and Sudarshan,
in which an early version of the Lindblad equation is derived, 
and has been re-discovered in the context of quantum process tomography \cite{CN97}, \cite{Al03}.
There an arbitrary superoperator ${\mathcal E}$ is written in the form
\begin{equation}\label{superGKS}
  {\mathcal E}(\rho) =\sum_{\alpha\beta} g^{\alpha\beta}\, F_\alpha\,\rho\,F_\beta^\ast
\end{equation}
implicitly defining the GKS matrix $\left( g^{\alpha\beta}\right)$.
For the special case $F_\alpha=|j\rangle\langle i|$, the GKS matrix is reduced to the Choi matrix
and thus inherits the criterion ``$ {\mathcal E}$ is completely positive iff  $g\ge 0$ ".

In the second part of this work, we investigated consequences for the time evolution of
general open quantum systems $S$ with environment $E$. Since this time evolution is completely positive,
the corresponding time-dependent GKS matrix $g(t)$ must be positive semi-definite.
This can also be confirmed directly by using a general Hamiltonian operator
for the time evolution of the total system $S+E$, but only up to order $t^2$ due to the complexity of the calculations.
Investigations of the dynamics of open quantum systems beyond the Markov approximation are 
part of current research, see for example \cite{Setal25}. They are mostly carried out 
within the framework of the time-dependent Lindblad equation. However, 
there is a certain difference between the time-dependent Lindblad equation 
and the GKS equation for $g(t)$, although both equations are equivalent, 
as we have shown in subsection \ref{sec:TDLE}.
The role of the GKS matrix for further physical applications of completely positive transformations,
such as in the theory of the measurement process must be reserved for future investigations.

\section*{Acknowledgment}\label{A}
I would like to thank Kyung Hoon Han for valuable comments on earlier versions of this paper,
and Frederik vom Ende for a literature reference.
I am also grateful to Vittorio Gorini for first-hand information on the early history of the GKSL equation.


\end{document}